\documentclass[12pt]{article}
\usepackage{mathrsfs}
\usepackage{amsmath}
\usepackage{mathrsfs}
\usepackage{amssymb}
\usepackage{color}
\usepackage{psfrag}
\usepackage{graphicx}
\usepackage{subfigure}
\usepackage{rotating}
\usepackage{cite}
\usepackage[colorlinks=true,linktocpage=true,linkcolor=blue,citecolor=blue]{hyperref}
\usepackage[section]{placeins}
\usepackage{amsfonts}
\usepackage{ifpdf}
\ifpdf
\else

\newcommand{\bea}{\begin{eqnarray}}
\newcommand{\eea}{\end{eqnarray}}
\newcommand{\be}{\begin{equation}}
\newcommand{\ee}{\end{equation}}
\renewcommand{\a}{\alpha}
\renewcommand{\b}{\beta}

\newcommand{\G}{\Gamma}

\newcommand{\la}{\lambda}
\newcommand{\La}{\Lambda}

\newcommand{\Om}{\Omega}
\newcommand{\half}{\frac{1}{2}}
\newcommand{\pa}{\partial}
\newcommand{\si}{\sigma}
\newcommand{\Si}{\Sigma}
\renewcommand{\t}{\theta}
\newcommand{\nn}{\nonumber}
\newcommand{\lan}{\langle}
\newcommand{\ran}{\rangle}
\newcommand{\z}{\mathcal}
\newcommand{\qq}{$Q$-$\bar{Q}$ }
\newcommand{\vs}[1]{\vspace{#1 mm}}

\begin{document}
\topmargin 0pt
\oddsidemargin 0mm

\vspace{2mm}

\begin{center}

{\Large \bf {Holographic quark-antiquark potential in hot, anisotropic Yang-Mills plasma}}

\vs{10}

{Somdeb Chakraborty\footnote{E-mail: somdeb.chakraborty@saha.ac.in}and Najmul Haque\footnote{E-mail: najmul.haque@saha.ac.in}}

 \vspace{4mm}

{\em

 Saha Institute of Nuclear Physics,
 1/AF Bidhannagar, Kolkata-700 064, India\\}

\end{center}

\vs{10}

\begin{abstract}
Using the gauge/gravity duality we calculate the heavy quark-antiquark potential in a hot, anisotropic and strongly coupled Yang-Mills plasma in (3+1)-dimensions. As the anisotropic medium we take a deformed version of $\z{N}=4$ super Yang-Mills theory at finite temperature following a recent work where the dual type IIB supergravity solution is also proposed. We turn on a small value of the anisotropy parameter, for which the gravity dual is known analytically (perturbatively), and compute the velocity-dependent quark-antiquark interaction potential when the pair is moving through the plasma with a velocity $v$. By setting $v = 0$ we recover the static quark-antiquark potential. We numerically study how the potential is modified in the presence of anisotropy. We further show numerically how the quark-antiquark separation (both in the static and the velocity-dependent case) and hence, the screening length gets modified by anisotropy. We discuss various cases depending upon the direction of the dipole and the 
direction of its propagation and make a comparative study of these cases. We are also able to obtain an analytical expression for the screening length of the dipole moving in a hot, anisotropic plasma in a special case.
\end{abstract}

\newpage
\tableofcontents
\section{Introduction} \label{intro}
Ever since the seminal work of Maldacena \cite{Mal}, further elaborated in \cite{Wit,Gub}, the AdS/CFT correspondence (see \cite{Aha} for a comprehensive review) and its subsequent generalizations have proved to be an indispensable tool for exploring the strongly coupled regime of large $N$ gauge theories (where $N$ is the rank of the gauge group). The correspondence, in its primitive incarnation, advocated the duality between  type IIB string theory on $AdS_{5} \times S_{5}$ (where $AdS$ stands for anti-de Sitter space) and $\z{N}=4, SU(N)$ super Yang-Mills (SYM) theory living on the $4$-dimensional boundary of $AdS_{5}$. Since then, it has been generalized to encompass a wider variety of gauge theories with known gravity duals and is now better called the gauge/gravity duality. One of the frontier areas where the duality has been particularly useful is the physics of quark-gluon plasma (QGP).
\\
Experiments at the Relativistic Heavy Ion Collider (RHIC) \cite{STAR} have provided various fascinating insights into the properties of quantum chromodynamics (QCD) matter at extreme high temperature, where it appears in the form of QGP. Most notably, the QGP does not behave as a weakly coupled gas consisting of quarks and gluons (as it should at very high temperature). Rather, there are strong indications that it resembles a strongly coupled fluid at the energy scale reached at RHIC \cite{Shu}. This  makes the theoretical description of QGP a challenging task. The strong coupling imposes severe limitations on the applicability of standard perturbative QCD techniques. Lattice  field theory is a powerful non-perturbative tool for exploring the static properties of strongly coupled gauge theories but  has limitations when called upon to explain various real-time properties that are relevant to QGP. In such a scenario, the gauge/gravity duality has emerged as a promising tool for exploring such strongly coupled 
non-Abelian plasma (\cite{Raj} provides a review of the applications of gauge/gravity duality to QGP).  Although the exact gravity dual to QCD has still eluded us, and the gauge theories and their gravity duals used for performing calculations are different from real world QCD, the results obtained so far have been quite encouraging. In fact, many of the results obtained exhibit a kind of universality among the different theories, the most notable among them being the celebrated $\eta/s$ ratio \cite{Pol,Kov}, where $\eta$ is the shear viscosity of the strongly coupled fluid and $s$ is its entropy density. By now there is a large body of literature which attempts at calculating various quantities of interest, like the drag force, the jet quenching parameter, the static and the velocity-dependent quark-antiquark potentials, the screening length, etc. in various QCD-like gauge theories in the deconfined phase using this duality. In many cases, the computed quantities have been in good qualitative agreement with 
their corresponding counterparts extracted from the experimental data. However, most of these works concern QGP which is locally isotropic. But, QGP, just after its creation in heavy ion collisions, is locally anisotropic and far away from equilibrium for a time $t < \tau_{out}$. Further, it becomes locally isotropic only after time $\tau_{iso} > \tau_{out}$, so that the standard hydrodynamic description of the plasma  makes sense only if we want to probe the plasma at time scale $t>\tau_{iso}$. One would, of course, like to make progress and study the plasma in the time scale $t<\tau_{out}$ when it is far away from equilibrium (a recent attempt towards this direction has been made in \cite{Che}). However, there lies an intermediate regime $\tau_{out}<t< \tau_{iso}$, where the plasma is in equilibrium but not in an isotropic state. To probe the QGP in this time window, it is imperative that one takes into account the inherent anisotropy. It has been proposed \cite{ani,ani2,ani3,ani4,ani5,ani6,ani7,ani8,ani9} 
that an inherently anisotropic hydrodynamic description, which involves a derivative expansion around an anisotropic state, can be used to study the plasma in this regime. In this time domain the plasma has unequal pressures in the longitudinal and the transverse directions leading to an anisotropic expansion of the plasma. While in reality the degree of anisotropy will decrease with time, here we shall always treat that it is independent of time over a suitable time scale. Attempts to investigate the anisotropic plasma in the framework of the gauge/gravity duality include \cite{MalRus,Ali,Har,Cai,Cai2,GubPuf,Amm,Bas,Amm2,Amm3,Nat,Erd,Kar,Alb,Erd2,Alb2,Hok,Hok2,Jan,Reb,Chak,Chak1}. Recently, Mateos and Trancanelli \cite{Mat,Mat2} proposed a completely regular type IIB supergravity solution dual to anisotropic plasma. Further work involving this particular gravity dual can be found in \cite{Cher,Gia,Cher2,Reb2,Fad,Mul,Cher3,Tra,Mamo} where the drag force, the jet quenching parameter, the stopping distance, 
the screening length, etc, were found out. In \cite{Reb3} it was shown that the $\eta/s$ ratio violates the conjectured bound in anisotropic plasma. It is thus natural that one should further extend this program to explore the consequences of the presence of anisotropy and see how it affects the various quantities of interest. Motivated by this, in this paper we address the issue of how the heavy quark-antiquark ($Q$-$\bar{Q}$) potential is modified in the presence of anisotropy.  We take the gravity dual as proposed in \cite{Mat,Mat2}. The static  \qq potential and the screening length were already considered in \cite{Gia} for such a system. Here we extend the analysis to the velocity-dependent case by considering  a heavy \qq pair moving through the plasma with a velocity $v$. While we have not restricted the value to be taken by $v$, we consider only small values of the anisotropy parameter, in which case the dual gravity solution is known perturbatively. Following \cite{Mal2,Rey,Rey2,Bra}, we employ the 
gauge/gravity duality to first compute the expectation values of certain Wilson loops, which are nonperturbative objects in gauge theories. In the background given in \cite{Mat}, we introduce a fundamental string probe and extremize the Nambu-Goto string world-sheet action in a static gauge. This, in turn, yields the expectation value of the Wilson loop, where the loop is the boundary of the above minimal area. For the dipole velocity $v<1$, the Wilson loop obtained is time-like and its expectation value can be related to the velocity-dependent interaction potential $V$ of the dipole using the prescription proposed in \cite{Liu}\footnote{Similar calculations were also carried out in \cite{Gui}.}. We plot the potential $V$ against the \qq separation $L$ for various values of the velocity $v$ and the anisotropy parameter $\tilde{a}$ and study how the introduction of a small anisotropy influences the potential. Unlike in the static case (as described in \cite{Gia}) where there were only two possible 
configurations of the dipole, here we shall see that the introduction of the velocity parameter gives rise to a plethora of possibilities, which we shall discuss one by one. We further probe the effect of anisotropy on the  \qq separation and consequently the screening length $L_{max}$. We are also able to obtain an analytic expression for the screening length in the anisotropic plasma in a special case. Although the static \qq potentials have already been provided in \cite{Gia}, we reproduce them here since they are recovered naturally in the $v=0$ limit of our analysis and nicely complement our results for the static \qq separation. In \cite{Cher3} the authors give an in-depth analysis of the screening length when the infinetely massive \qq pair (or the heavy meson or quarkonium, as they call it) moves in hot, anisotropic plasma. In this paper, one can read off the screening length from the plot of the \qq separation. Wherever the results overlap, they are in agreement with those obtained in \cite{Cher3}.\
\
The paper is organized as follows. In section \ref{dual}, we briefly review the model and the dual geometry and discuss the general set-up. Section \ref{pot} is the core of the paper where we compute the \qq separation and the potential and provide the numerical results. We also calculate the screening length analytically in a special case. Section \ref{pot} is divided into five subsections corresponding to the different cases we consider.  In section \ref{comp}, we compare our results  for the  different cases considered and  in section \ref{disc} follow it up with comparison with some other models available in the literature. Finally, in section \ref{conc} we summarize our work and conclude.

\section{The dual geometry} \label{dual}
In this section we briefly review the gravity dual of the gauge theory we are interested in and discuss the general set-up of the problem. We will take the gauge theory as a deformed version of $\z{N}=4$, SU(N) YM theory ($N$ being the number of colors) at large t'Hooft coupling $\la=g_{YM}^{2}N$ where the deformation is achieved by introducing a $\t$-term in the action as
\be 
S=S_{SYM}+\frac{1}{8 \pi^{2}}\int \t(x^3) \text{Tr}  F \wedge F
\ee
where $\t(x^3) \propto x^{3}$ (we take $(t,x^{1},x^{2},x^{3})$ as the gauge theory coordinates). The presence of the non-zero $\t$-term breaks the $SO(3)$ rotational symmetry down to a $SO(2)$ symmetry in the $x^{1}$-$x^{2}$ plane and makes the theory anisotropic. In the context of heavy ion collisions, $x^{3}$ will correspond to the direction of beam whereas the $x^{1},x^{2}$-directions span the transverse plane. The dual gravity background was given in \cite{Mat,Mat2} inspired by \cite{Aze} and reads in the string frame,
\be \label{MT}
ds^{2}=r^{2}\left(-\z{FB}dt^{2}+(dx^{1})^{2}+(dx^{2})^{2}+\z{H}(dx^{3})^{2}+\frac{dr^{2}}{r^{4}\z{F}}\right)+e^{\half \phi}d\Om_{5}^{2},
\ee
\be 
\chi=ax^{3}, \qquad \phi=\phi(r)
\ee
where $\chi$ is the axion, proportional to the anisotropic direction $x^{3}$, the proportionality constant $a$ being the anisotropy parameter and $\phi$ is the dilaton. $r$ is the $AdS$ radial coordinate with the boundary at $r=\infty$ and the horizon at $r=r_{h}$. Here  $d\Om_{5}^{2}$ is the metric on  the five-sphere $S_{5}$ and we have set  the common radius of the $AdS$ space and $S_{5}$  to unity. There is also a RR self-dual five-form which will not play any role in our discussion here.  Anisotropy is introduced through the axion, the dual to the gauge theory $\t$-term. The anisotropy parameter $a$ turns out to be \cite{Mat} $a=\la n_{D7}/4 \pi N$ where $n_{D7}$ is the density of $D7$-branes (which acts as the magnetic source of the axion) along the $x^{3}$-direction. The $D7$-branes wrap around $S_{5}$ and extend along the transverse directions, $x^{1},x^{2}$. Thus in the gravity dual the presence of anisotropy can be attributed to the existence of anisotropic extended objects. Note that the $D7$-
branes do not extend along the radial direction. Hence, they do not reach the boundary and do not contribute any new degrees of freedom to the theory. $\z{F,B,H}$  are all functions of the radial coordinate $r$ and are known analytically only in the limiting cases when the temperature is very high or low. Otherwise, they are known numerically in the intermediate range. $\z{F}$ is the `blackening factor' which vanishes at the horizon, i.e., $\z{F}(r_{h})=0$. The degree of anisotropy can be controlled by tuning the parameter $a$. In this paper, we shall be concerned with weakly anisotropic plasma (the small $a$ or high temperature $T$ limit, such that $a/T \ll 1$) in which case the functions $\z{F,B,H}$ can be expanded to leading order in $a$ around the black $D3$-brane solution,
\bea
\z{F}(y)&=&1-\frac{1}{y^{4}}+a^{2}\z{F}_{2}(y)+\z{O}(a^{4}),\nn \\
\z{B}(y)&=&1+a^{2}\z{B}_{2}(y)+\z{O}(a^{4}),\nn \\
\z{H}(y)&=&e^{-\phi(y)} \qquad \text{with} \qquad \phi(y)=a^{2}\phi_{2}(y)+\z{O}(a^{4})
\eea
where
\bea
\z{F}_{2}(y)&=&\frac{1}{24r_{h}^{2}y^{4}}\left[8(y^{2}-1)-10\log2 +(3y^{4}+7)\log\left(1+\frac{1}{y^{2}}\right) \right], \nn \\
\z{B}_{2}(y)&=&-\frac{1}{24 r _{h}^{2}}\left[\frac{10}{1+y^{2}}+\log\left(1+\frac{1}{y^{2}}\right) \right], \nn \\
\phi_{2}(y)&=&-\frac{1}{4r_{h}^{2}}\log\left(1+\frac{1}{y^{2}}\right)
\eea
and we have defined the dimensionless quantity $y=r/r_{h}$. The temperature is given by
\be 
T=\frac{r_{h}}{\pi }+\frac{a^{2}}{r_{h}}\frac{(5\log2-2)}{48\pi}+\z{O}(a^{4})
\ee
which can be inverted to yield the horizon position in terms of the temperature, which, in the limit $a/T \ll 1$, reads
\be  \label{temp}
r_{h} \sim \pi T\left[ 1-a^{2}\frac{5\log2-2}{48\pi^{2}T^{2}}\right].
\ee
We introduce a fundamental string in this background and evaluate the Nambu-Goto action $S$. By extremizing this action we find the expectation value of the relevant Wilson loop. Assuming the string to move along $x^{i}$ with a velocity $v$ and the string endpoints to lie along $x^{j}$, separated by a distance $L$ (which in the dual gauge theory translates to the quark-antiquark separation), the Wilson loop so formed is a rectangle with a short side $L$ along $x^{j}$ and a long side $\z{T}$ along any time-like direction in the $t$-$x^{i}$ plane. We further assume that $L \ll \z{T}$ to ensure that the string world-sheet remains invariant under time translations. The expectation value of the Wilson loop so formed is,
\be  \label{W1}
\lan W\ran=e^{i\left(S-S_{0} \right)}
\ee
where $S_{0}$, the Nambu-Goto action corresponding to two disjoint strings (dual to a non-interacting quark and antiquark) is subtracted to offset the divergence in $S$. Now the \qq interaction potential $V(L)$ is extracted from a knowledge of the expectation value of the Wilson loop via the working definition,
\be \label{W2}
\lan W\ran=e^{iV(L)\z{T} }.
\ee
For the static \qq separation and potential one needs to consider only two possibilities: the dipole lying along the anisotropic direction $x^{3}$ or in the transverse plane. However, the introduction of the velocity $v=\tanh \eta$ ($\eta$ is the rapidity parameter) opens up the following possibilities:

\begin{enumerate}{}{}
\item Motion in transverse plane, dipole lies perpendicular to direction of motion in the plane.
\item Motion in transverse plane, dipole along $x^{3}$.
\item Motion along $x^{3}$, dipole in transverse plane.
\item Motion in transverse plane, dipole parallel to direction of motion in the plane.
\item Motion and dipole, both along the anisotropic direction.\footnote{Of course, there exist other possibilities where the dipole can have any arbitrary orientation with respect to its direction of motion, which, itself, can be in any arbitrary direction. However, we do not consider these cases here.}
\end{enumerate}

\section{\qq separation and \qq potential} \label{pot}
In this section we discuss the different cases, alluded to above, one in each subsection,  and for each case we numerically study the \qq separation with varying values of the rapidity parameter $\eta$ and the anisotropy parameter $a$ and see how the separation and hence, the screening length gets affected when we turn on a small value of $a$. We also compute the \qq potential (both velocity-dependent and static) and observe the modifications brought about by anisotropy. Further, we provide an analytic expression for the screening length in a special case.

\subsection{Motion in transverse plane, dipole lies perpendicular to direction of motion in the plane} \label{12}
In this case the motion is wholly contained in the transverse plane and the dipole presents itself perpendicular to the direction of its motion. We first set our axes such that the dipole moves along $x^{1}$ while itself being aligned along $x^{2}$. Then we go to the rest-frame $(t',x^{1\prime})$ of the \qq pair via the following coordinate transformation,
 \be 
 \begin{aligned}
 & dt=\cosh \eta dt'-\sinh \eta dx^{1\prime},\\
 & dx^{1}=-\sinh \eta dt'+\cosh \eta dx^{1\prime}.
 \end{aligned}
 \ee
Now the \qq pair and hence the Wilson loop can be regarded as static in a plasma that is moving with a velocity $v$ in the negative $x^{1\prime}$ direction. This implies that the rectangular Wilson loop spans the $t'$ (since $x^{1\prime}$ is fixed in this rest-frame) and $x^{2}$ directions with sides $\z{T}$ and $L$ respectively. In terms of the boosted coordinates the metric (\ref{MT}) can be rewritten as
\bea \label{MTboosted}
ds^{2}\!\!\!\!&=&\!\!\!\!-A(r)dt^{2}-2B(r)dtdx^{1}+C(r)(dx^{1})^{2}+r^{2}\left((dx^{2})^{2}+\z{H}(dx^{3})^{2} +\frac{dr^{2}}{r^{4}\z{F}}\right)+e^{\half \phi}d\Om_{5}^{2} \nn\\
&=&G_{\mu \nu}dx^{\mu}dx^{\nu}
\eea
where
\bea
A(y)&=&(yr_{h})^{2}\left[ 1-\frac{\cosh^{2}\eta}{y^{4}} +a^{2}\cosh^{2}\eta \left\{\z{F}_{2}+\z{B}_{2}\left(1-\frac{1}{y^{4}}\right) \right\}\right],\nn \\
B(y)&=&(yr_{h})^{2}\sinh \eta \cosh \eta \left[\frac{1}{y^{4}}-a^{2} \left\{\z{F}_{2}+\z{B}_{2}\left(1-\frac{1}{y^{4}}\right) \right\}\right],\nn \\
C(y)&=&(yr_{h})^{2}\left[ 1+\frac{\sinh^{2}\eta}{y^{4}} -a^{2}\sinh^{2}\eta \left\{\z{F}_{2}+\z{B}_{2}\left(1-\frac{1}{y^{4}}\right) \right\}\right].
\eea
(Note that since we shall be using the primed coordinates from now on, we have got rid of the primes for simplicity. Also, we have suppressed the $y$-dependence of the quantities $\z{F}_{2},\z{B}_{2}$. Further, we have expressed $A,B$ and $C$ as functions of the scaled radial coordinate $y$.) In this background we evaluate the Nambu-Goto string world-sheet action,
\be \label{NG}
S=\frac{1}{2\pi \a'}\int d\si d\tau \sqrt{-\text{det}g_{\a \b}}
\ee
where $g_{\a \b}$ is the induced metric on the string world-sheet,
\be 
g_{\a \b}=G_{\mu \nu}\frac{\pa x^{\mu}}{\pa \xi^{\a}}\frac{\pa x^{\nu}}{\pa \xi^{\b}}.
\ee
Here $G_{\mu \nu}$ is the ten-dimensional metric as given in (\ref{MTboosted}) and $\xi^{\a,\b}$ are the world-sheet coordinates, $\xi^{0}=\tau$ and $\xi^{1}=\si$. We choose the static gauge for evaluating (\ref{NG}) as $\tau=t,\si=x^{2}$ where $-L/2 \leq x^{2}\leq +L/2$ and $r=r(\si), x^{1}(\si)=x^{3}(\si)=\text{constant}$. We wish to determine the string embedding $r(\si)$ supplemented by the boundary condition $r\left(x^{2}=\pm L/2\right) \rightarrow \infty$. Equipped with the above parametrization, the Nambu-Goto action (\ref{NG}) can be written as,
\be 
S=\frac{\z{T}}{2\pi \a '}\int\limits_{-L/2}^{L/2}d\si \sqrt{A\left(G_{22}+G_{rr}(\pa_{\si}r)^{2}\right)}.
\ee
At this stage, for convenience, let us define the new dimensionless quantities, $\tilde{\si}=\si/r_{h}$ and $l=L/r_{h}$, the scaled \qq separation, whence we can rewrite the above action as
\be \label{NG1}
S=\frac{\z{T}r_{h}}{2\pi \a '}\int\limits_{-l/2}^{l/2}d\tilde{\si}\z{L}
\ee
where 
\be 
\z{L}=\sqrt{A\left(G_{22}+G_{rr}y'^{2}\right)}
\ee
is the Lagrangian density and $\pa_{\tilde{\si}}y=y'$. Note that $\z{L}$ does not have any explicit $\tilde{\si}$-dependence which at once allows us to extract the conserved quantity,
\be  \label{constant12}
\z{L}-y'\frac{\pa \z{L}}{\pa y'}=\frac{AG_{22}}{\sqrt{A(G_{22}+G_{rr}y'^{2})}}=K
\ee
which, in turn, yields,
\be 
y'=\frac{1}{K}\sqrt{\frac{G_{22}}{G_{rr}}}\sqrt{AG_{22}-K^{2}}.
\ee
Upon integration we obtain
\be 
l=2\int\limits_{0}^{l/2}d\tilde{\si}=2K\int\limits_{y_{t}}^{\infty}dy \sqrt{\frac{G_{rr}}{G_{22}}}\frac{1}{\sqrt{AG_{22}-K^{2}}}.
\ee
The limits in the second integration require a little explanation. Recall that $y$ is the scaled radial coordinate and the string hangs down starting from $y=\infty$ (where the boundary gauge theory lives) up to $y_{t}$ (which we shall find shortly), where it turns back and rises again up to $y=\infty$. Plugging in the explicit expressions for the metric components, we arrive at,
\be 
l=\frac{2\tilde{K}}{r_{h}^{2}}\int\limits_{y_{t}}^{\infty}dy \frac{1}{\sqrt{\left(y^{4}-1+\frac{\tilde{a}^{2}}{24}\Si(y)\right)\left(y^{4}-y_{c}^{4}+\frac{\tilde{a}^{2}}{24}\La(y)\cosh^{2}\eta\right)}}
\ee
where we have defined,
\be 
\Si(y)=8(y^{2}-1)-10\log2+(3y^{4}+7)\log\left(1+\frac{1}{y^{2}}\right),
\ee
\be 
\La(y)=2(1-y^{2})-10\log2+2(y^{4}+4)\log\left(1+\frac{1}{y^{2}}\right),
\ee
and $\tilde{a}=a/r_{h} (\sim a/\pi T), \tilde{K}=K/r_{h}^{2}, y_{c}^{4}=\cosh^{2}\eta+\tilde{K}^{2}$. Using (\ref{temp}) one can now find the actual \qq separation as
\be  \label{L12}
L=\frac{2\tilde{K}}{\pi T}\left(1+\frac{\tilde{a}^{2}}{48}(5\log2-2) \right)\int\limits_{y_{t}}^{\infty}dy \frac{1}{\sqrt{\left(y^{4}-1+\frac{\tilde{a}^{2}}{24}\Si(y)\right)\left(y^{4}-y_{c}^{4}+\frac{\tilde{a}^{2}}{24}\La(y)\cosh^{2}\eta\right)}}.
\ee
As mentioned earlier, to perform the integration, one needs to specify $y_{t}$. The turning point is found out by demanding that the terms in the denominator vanish separately (which is equivalent to demanding that $y'$ vanishes at these points) and accepting the larger one among them. As one can easily verify, the first term in the denominator vanishes at $y=1$, since $\Si(1)=0$, thereby,  furnishing a turning point at $y_{t1}=1$ up to $\z{O}(a^{2})$. To find the turning point $y_{t2}$ arising from the second term, we assume the anisotropy parameter $\tilde{a}$\footnote{Since in our analysis $a$ always appears in the form $a^{2}/r_{h}^{2} \equiv \tilde{a}^{2}$, we shall, henceforth, call $\tilde{a}$ the anisotropy parameter.} to be small and we need to find a solution to
\be 
y_{t2}^{4}-y_{c}^{4}+\frac{\tilde{a}^{2}}{24}\La(y_{c})\cosh^{2}\eta=0.
\ee
Note that we have evaluated $\La$ at $y=y_{c}$ since the term is already at $\z{O}(\tilde{a}^{2})$ and consequently, the error incurred is $\sim \z{O}(\tilde{a}^{4})$. This has a solution
\be 
y_{t2}=y_{c}\left(1-\frac{\tilde{a}^{2}}{24y_{c}^{4}}\La(y_{c})\cosh^{2}\eta \right)^{1/4}.
\ee
It can be shown that $y_{t2}>1$ always, so that we take it to be the actual turning point $y_{t}$. As expected, by setting $\tilde{a}=0$ we recover the turning point $y_{c}$ in the isotropic case. Now (\ref{L12}) gives the \qq separation as a function of the constant $\tilde{K}$. However, it is not possible to perform the integration analytically. Hence, we shall resort to numerical integration and show how the \qq separation is affected by the presence of anisotropy for various values of the rapidity parameter $\eta$ and the anisotropy parameter $\tilde{a}$. Further, by setting $\eta=0$ we obtain the static \qq separation. We postpone the discussion of our numerical results till we give the \qq potential.\\
Changing the integration variable from $\tilde{\si}$ to $y$ we can rewrite the action (\ref{NG1}) as
\be 
S=\frac{\z{T}r_{h}}{\pi \a '}\int\limits_{y_{t}}^{\infty}dyA\sqrt{\frac{G_{22}G_{rr}}{AG_{22}-K^{2}}}.
\ee
Putting the explicit expression for the metric components one finally has
\bea \label{S12}
S&=&\frac{\z{T}r_{h}}{\pi \a '}\int\limits_{y_{t}}^{\infty}dy\frac{y^{4}-\cosh^{2}\eta+\frac{\tilde{a}^{2}}{24}\La(y)\cosh^{2}\eta}{\sqrt{\left( y^{4}-1+\frac{\tilde{a}^{2}}{24}\Si(y)\right)\left(y^{4}-y_{c}^{4}+\frac{\tilde{a}^{2}}{24}\La(y)\cosh^{2}\eta \right)}} \nn\\
&\equiv&\frac{\z{T}r_{h}}{\pi \a '}\int\limits_{y_{t}}^{\infty}dy\z{S}^{ani}.
\eea
A mere inspection reveals that the integral, as it stands, diverges. This is because the action $S$ contains the self-energy contributions from the free  heavy \qq pair which, themselves, are divergent. To obtain the \qq interaction potential $V(L)$ we need to cure this divergence, which is done by subtracting from $S$ the action $S_{0}$ of a free \qq pair whence from (\ref{W1},\ref{W2})
\be \label{V}
V(L)=\frac{S-S_{0}}{\z{T}}.
\ee
To compute $S_{0}$ we consider an open string hanging down the radial direction (in gauge theory it translates to a single quark/antiquark propagating in the same background as before) and employ the static gauge condition, $\tau=t, \si=r, x^{1}=x^{1}(\si)$ and $x^{2},x^{3}$ are independent of $\tau,\si$. We evaluate the Nambu-Goto action (and multiply by two to take into account the contribution from the quark and the antiquark) which takes the form,
\bea
S_{0}&=&\frac{\z{T}}{\pi \a '}\int\limits_{r_{0}}^{\infty} dr \sqrt{AG_{rr}+(x^{1\prime})^{2}(AC+B^{2})} \nn \\
&\equiv&\frac{\z{T}}{\pi \a '}\int\limits_{r_{0}}^{\infty} dr \z{L}_{0}.
\eea
As before, $S_{0}$ too does not have any explicit $x^{1}$-dependence implying that there exists a conserved quantity,
\be
\frac{\pa \z{L}_{0}}{\pa x^{1\prime}}=\left(AC+B^{2}\right)\frac{x^{1\prime}}{\z{L}_{0}}=\text{constant}=K_{0}
\ee
which yields,
\bea \label{x'12}
(x^{1\prime})^{2}&=&K_{0}^{2}\frac{AG_{rr}}{(AC+B^{2})(AC+B^{2}-K_{0}^{2})} \nn \\
&=&\frac{\tilde{K}^{2}_{0}}{r_{h}^{4}}\frac{\left(y^{4}-\cosh^{2}\eta+\frac{\tilde{a}^{2}}{24}\La(y)\cosh^{2}\eta\right)}{\left(y^{4}-1+\frac{\tilde{a}^{2}}{24}\La(y)\right)\left( y^{4}-1+\frac{\tilde{a}^{2}}{24}\Si(y)\right)\left(y^{4}-1-\tilde{K}_{0}^{2}+\frac{\tilde{a}^{2}}{24}\La(y)\right) }
\eea
(we have used the scaling $\tilde{K}_{0}=K_{0}/r_{h}^{2}$). Note that in the expression for $S_{0}$ we have not specified the lower limit of the integration $r_{0}$ (or $y_{0}$ after scaling), which we shall now determine. For a string (corresponding to a free quark/antiquark) hanging down we expect it to extend all the way to the horizon at $y=1$. This is the case  when the string moves through the isotropic background. In particular, this implies that the string can not encounter a turning point before $y=1$. In our case, the possible turning points can be found out from (\ref{x'12}) by demanding  that $x^{1\prime}=\infty$ at those points. Now the first two terms in the denominator of (\ref{x'12}) give the turning point $y_{0}=1$ up to $\z{O}(\tilde{a}^{2})$ since $\Si(1)=\La(1)=0$. However, the third term (which contains the unspecified constant $\tilde{K}_{0}$) gives a turning point $y_{0}^{4} \sim 1+\tilde{K}_{0}^{2}+\z{O}(\tilde{a}^{2})$ which is greater than zero even for the isotropic case. Taking 
cue from the isotropic case we eliminate this possibility by constraining the value of $\tilde{K}_{0}$ such that the zero of this term coincides with the zero of the numerator. This at once provides us an expression for $\tilde{K}_{0}$ as
\be 
\tilde{K}_{0}^{2}=\sinh^{2}\eta\left( 1-\frac{\tilde{a}^{2}}{24}\La\left(y=\sqrt{\cosh \eta}\right)\right).
\ee
We can now recast the action as
\bea \label{S012}
S_{0}\!\!\!\!&=&\!\!\!\!\frac{\z{T}}{\pi \a '}\int\limits_{r_{h}}^{\infty} dr \sqrt{AG_{rr}}\sqrt{\frac{AC+B^{2}}{AC+B^{2}-K_{0}^{2}}}\nn \\
\!\!\!\!&=&\!\!\!\! \frac{\z{T}r_{h}}{\pi \a '}\int\limits_{1}^{\infty} dy\frac{\sqrt{\left(y^{4}-\cosh^{2}\eta+\frac{\tilde{a}^{2}}{24}\La(y)\cosh^{2}\eta\right)\left(y^{4}-1+\frac{\tilde{a}^{2}}{24}\La(y)\right)}}{\sqrt{\left( y^{4}-1+\frac{\tilde{a}^{2}}{24}\Si(y)\right) \left(y^{4}-\cosh^{2}\eta +\frac{\tilde{a}^{2}}{24}\left(\La(y)+\La(y=\sqrt{\cosh \eta})\sinh^{2}\eta\right)\right)}}\nn \\
\!\!\!\!&\equiv&\!\!\!\!\frac{\z{T}r_{h}}{\pi \a '}\int\limits_{1}^{\infty} dy \z{S}^{ani}_{0}.
\eea
Inserting (\ref{S12}) and (\ref{S012}) in (\ref{V}) and then using (\ref{temp}) we can now write
\be  \label{Vfin}
\frac{V}{T}=\sqrt{\la}\left( 1-\frac{\tilde{a}^{2}}{48}(5\log2-2)\right)\left(\int\limits_{y_{t}}^{\infty}dy\z{S}^{ani}-\int\limits_{1}^{\infty}dy\z{S}_{0}^{ani}\right)
\ee
where we have used the standard AdS/CFT dictionary $R^{4}=\la \a '^{2}$ (with $R$ set to unity here) to express our final result in terms of quantities pertaining to the gauge theory. Evaluating (\ref{Vfin}) involves performing integrals which can not be handled analytically. We, therefore, fall back upon numerical means to perform these integrals and numerically show our results. We compute the \qq separation $L$ for various values of the rapidity parameter $\eta$ and the anisotropy parameter $\tilde{a}$ as a function of the constant $\tilde{K}$, numerically invert (\ref{L12}) to express $\tilde{K}$ in terms of $L$ and plug it in (\ref{S12}) to finally obtain the \qq potential as a function of the \qq separation. Here we shall provide our numerical results for both the \qq separation and the \qq potential. In particular, by setting $\eta=0$ we recover the static \qq potential, which was already found out in \cite{Gia}.
\begin{figure}[t]
\begin{center}
\subfigure[]{
\includegraphics[width=7.4cm,height=7.4cm, angle=-0]{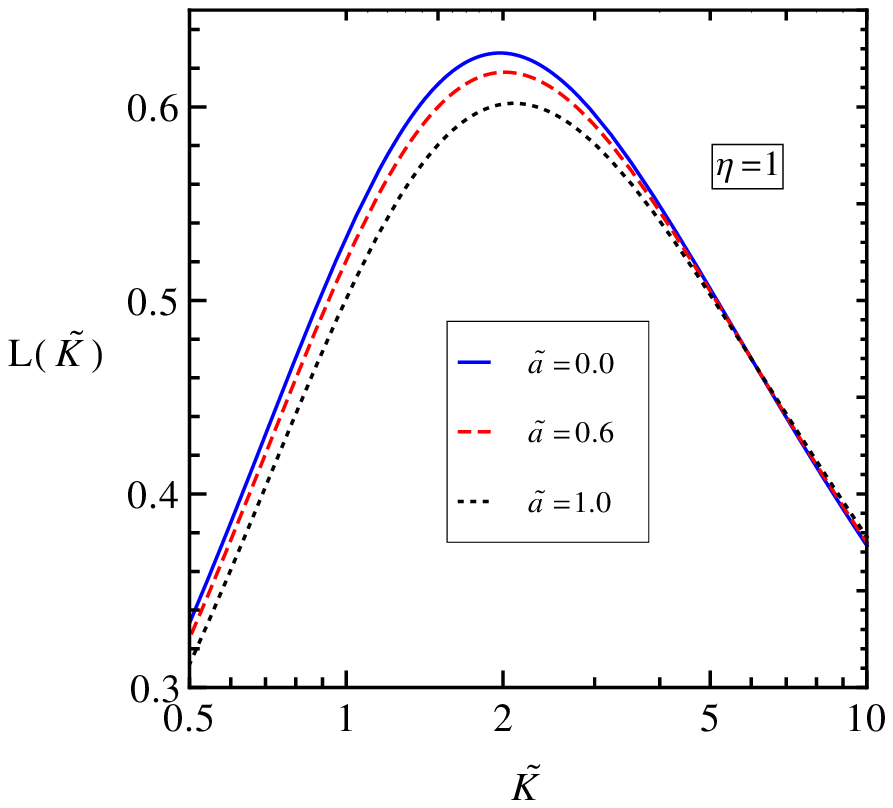}}
\hspace{5mm}
\subfigure[]{
\includegraphics[width=7.4cm,height=7.4cm, angle=-0]{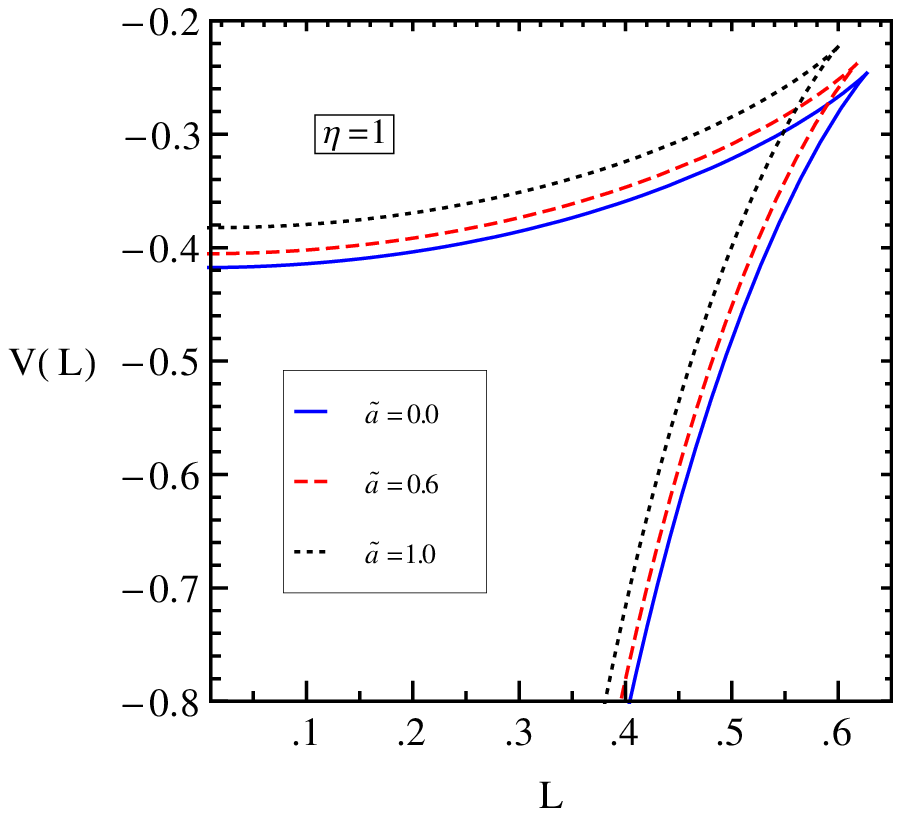}}
\caption{\label{12eta1} \small{(a) shows the plot of 
quark-antiquark separation $L$ (normalized) as a function of integration const. $\tilde{K}$ with $\eta=1$ for
different values of the anisotropy parameter $\tilde{a}$ when the dipole lies perpendicular to its velocity but both lie in the transverse plane. (b) shows the plot of properly normalized quark-antiquark potential $V$ as a function of $L$ with $\eta=1$ for the same set of anisotropy parameter values.}}
\end{center}
\end{figure}
\begin{figure}[t]
\begin{center}
\subfigure[]{
\includegraphics[width=7.6cm,height=7.6cm, angle=-0]{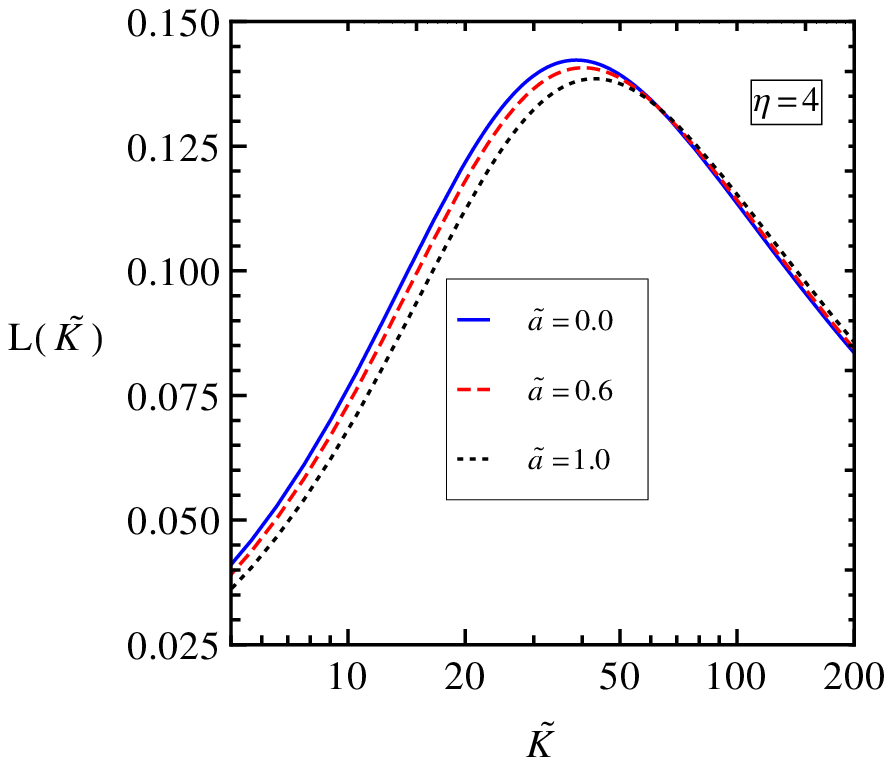}}
\hspace{5mm}
\subfigure[]{
\includegraphics[width=7.3cm,height=7.3cm, angle=-0]{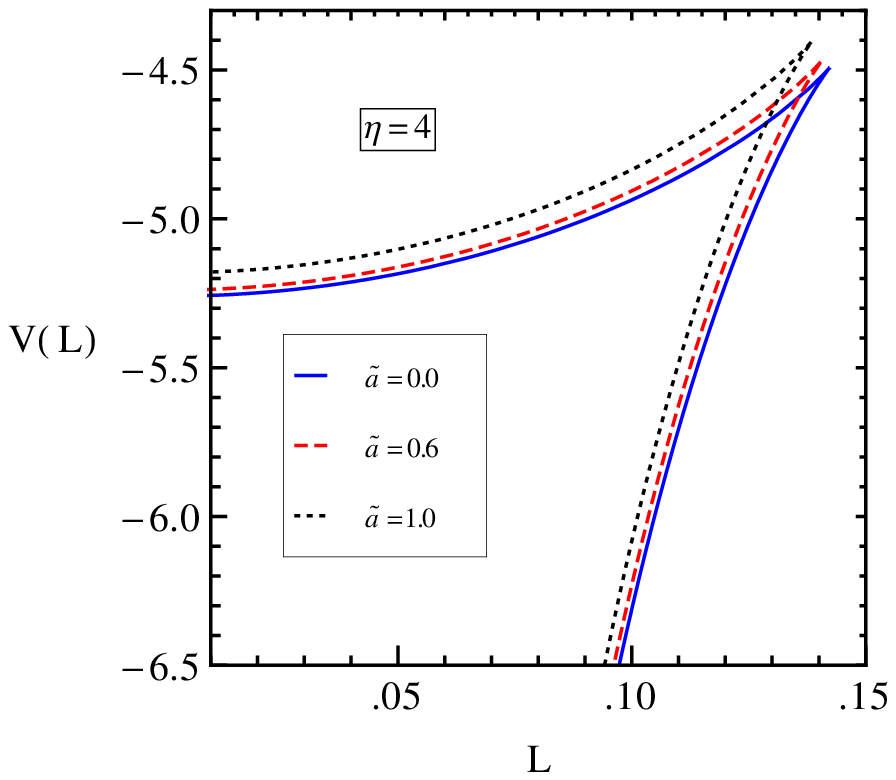}}
\caption{\label{12eta4} \small{(a) shows the plot of 
$L$ (normalized) as a function of $\tilde{K}$ with $\eta=4$ for
different values of $\tilde{a}$ when the dipole lies perpendicular to its velocity but both lie in the transverse plane. (b) shows the plot of properly normalized $V$ as a function of $L$ with $\eta=4$ for the same set of values of $\tilde{a}$.}}
\end{center}
\end{figure}
\begin{figure}[t]
\begin{center}
\subfigure[]{
\includegraphics[width=7.2cm,height=7.2cm, angle=-0]{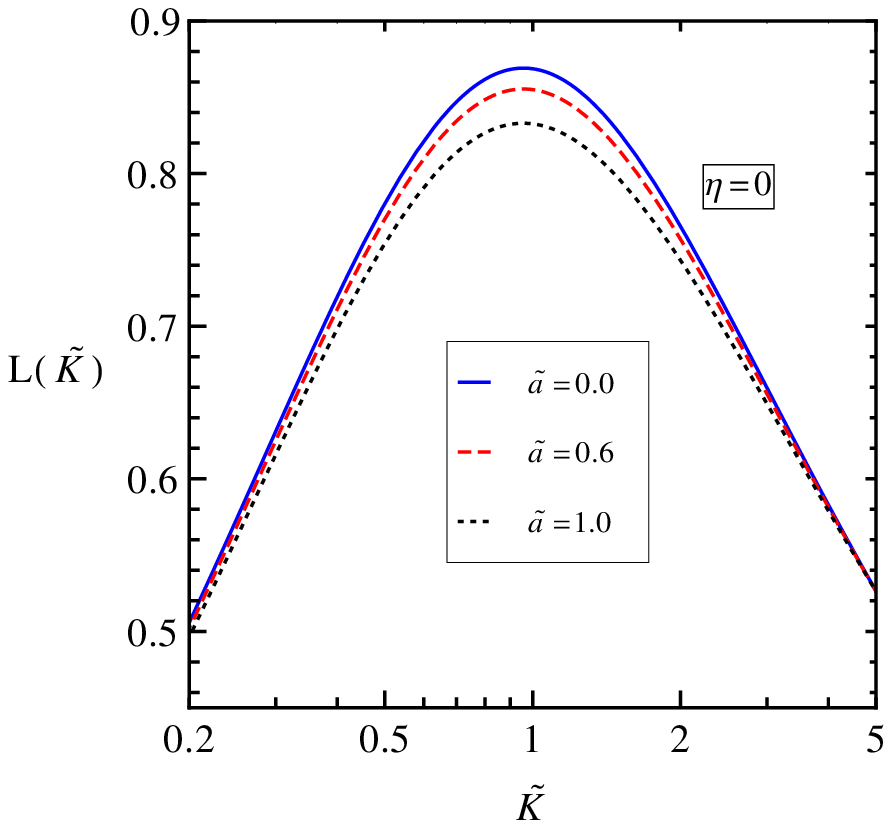}}
\hspace{5mm}
\subfigure[]{
\includegraphics[width=7.6cm,height=7.6cm, angle=-0]{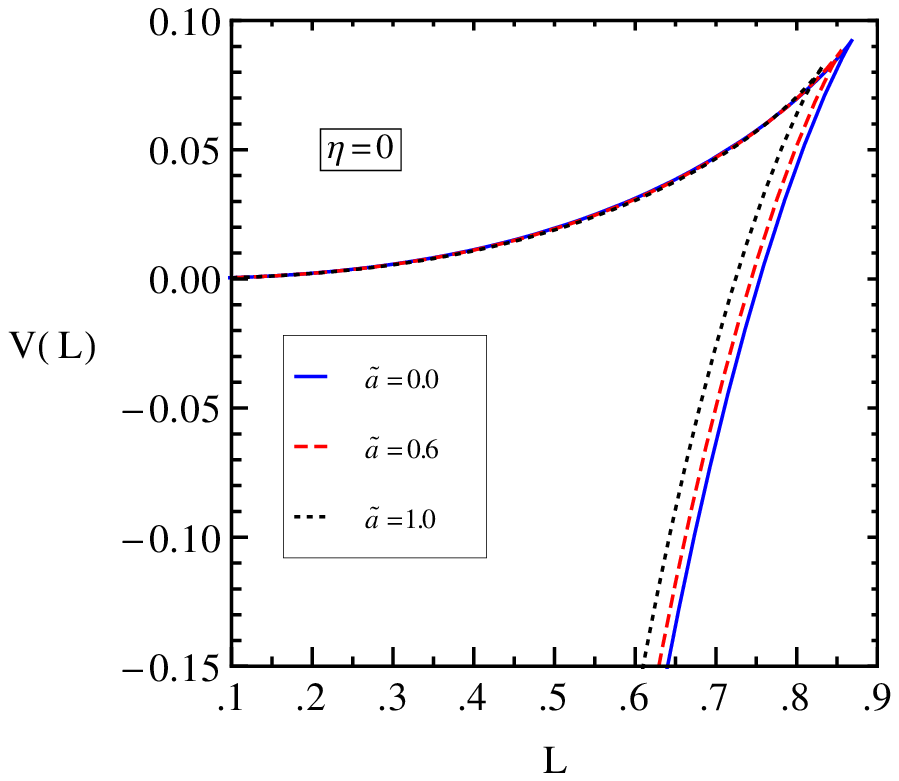}}
\caption{\label{12eta0} \small{(a) shows the plot of 
$L$ (normalized) as a function of $\tilde{K}$ with $\eta=0$ for
different values of $\tilde{a}$ when the dipole lies in the transverse plane. (b) shows the plot
of properly normalized $V$ as a function of $L$ with
$\eta=0$ for the same set of values of $\tilde{a}$.}}
\end{center}
\end{figure}
In Figs.\ref{12eta1} and \ref{12eta4} we have presented the  \qq separation $L(\tilde{K})$ as a function of $\tilde{K}$ and the \qq potential $V(L)$ as a function of $L$ for $\eta=1~ (v=0.762)$ and $\eta=4 ~(v=0.999)$ respectively. Each figure has two parts - part (a) showing the $L(\tilde{K})$-$\tilde{K}$ plot and part (b) showing the $V(L)$-$L$ plot. While the qualitative pattern of  the plots in both the figures are the same, the details differ. So we shall take Fig.\ref{12eta1} as the prototype case and discuss the results. First of all, we find from part (a) that as $\tilde{K}$ increases the separation $L(\tilde{K})$ increases till it reaches a maximum $L_{max}$ after which it again falls off. $L_{max}$ is interpreted as the screening length\footnote{Note that our definition of the screening length differs slightly from that used in \cite{Cher3}.} of the dipole, i.e., beyond this critical value of $L$ the screening effect of the plasma is sufficient to break the dipole. We observe that the effect of 
anisotropy is to suppress the screening length thereby encouraging the melting of the dipole. In particular, the degree of suppression of $L_{max}$ is more for stronger anisotropy. The deviation from the isotropic curve is more pronounced for lower $\tilde{K}$ (before $L_{max}$ is attained) than for higher $\tilde{K}$ (after $L_{max}$). For $L<L_{max}$ there can be two dipoles at a fixed $L$ for two different values of $\tilde{K}$. To understand at which one of the $\tilde{K}$ values the dipole will actually exist we need to analyze the $V(L)$-$L$ plot. The \qq potential has two branches corresponding to two different values of $\tilde{K}$. The upper branch corresponds to smaller value of $\tilde{K}$ whereas the lower branch corresponds to higher value of $\tilde{K}$.  Of course, the lower branch has lower energy and consequently, it is the preferred state of the dipole. So, even if a dipole is in the upper branch it will not be in a stable configuration and the dipole will make a transition to the lower 
branch. As we turn on a small anisotropy both the branches of the potential  shift slightly upwards. Since the upper branch is physically insignificant, corresponding to an unstable state, we shall confine our discussion to the lower branch only. The marginal upward shift in the potential indicates that the dipole is now loosely bound, the shift being more prominent for higher values of the anisotropy parameter $\tilde{a}$. This fits in with our conclusion from the $L(\tilde{K})$-$\tilde{K}$ plot that anisotropy enhances the screening effect of the medium. Further notice that in both  cases the potential is always negative. In Fig.\ref{12eta0} the plots for the static case are shown. Since the basic nature is the same, we shall not elaborate upon our results and briefly mention the salient features of the plots emphasizing the differences from the velocity-dependent cases. First of all, notice that unlike the moving dipole case, now the deviation from the isotropic curve in the $L(\tilde{K})$-$\tilde{K}$ 
plot is appreciable on either side of $L_{max}$.  Also note that now $L_{max}$ is much higher for the $\eta=0$ case and steadily decreases as we increase $\tilde{a}$.  In the $V(L)$-$L$ plot the lower branch suffers a small elevation whereas, the insignificant upper branch is largely insensitive to changes in $\tilde{a}$. The new feature  that now emerges is that the static potential crosses zero and becomes positive at a particular value $L=L_{p}$.
As is well-known, in the confined phase, the potential has two parts: $V(L) \sim -\frac{\a}{L}+\si L$. 
In the deconfined phase only the Coulomb part is modified (medium modification) whereas the string tension (confining) term goes to zero. However, it has been discussed in \cite{Peter,Stri,Patra} that in the deconfined phase
it is not sufficient to take only the screened Coulomb part of the potential. Rather, one must also take into account the medium depedent contribution arising from the string term. These two terms have opposite signs and at large separation the string term dominates over the Coulomb potential. Here $L_{p}$ denotes the separation beyond which the string tension term starts to dominate. This feature is nicely captured in the plot here in a qualitative manner.  In fact, there will be a critical velocity $v_{p}=\tanh \eta_{p}$ (whose value will, in general, also depend upon $\tilde{a}$)  beyond which the potential will not contain any positive piece.\\
We shall now obtain an analytical expression for the screening length, albeit in a special case. The \qq separation $L$ has already been given in (\ref{L12}). We have mentioned earlier that, in general, the integration appearing in (\ref{L12}) can not be done analytically. Of course, this is not to be thought of as the artifact of our anisotropic background. Rather, it is a handicap present in the isotropic case, too. Here, to facilitate analytical manipulation, we shall confine ourselves to the ultra-relativistic regime where $\eta$ is large, in which case the turning point $y_{t}$ also becomes very large but assume the product $\tilde{a}^{2}\cosh^{2}\eta$ is sufficiently small. In this special case the first term in the denominator in (\ref{L12}) lends itself to a binomial expansion. Here, for the sake of simplicity, we shall consider only the leading order term in the aforesaid expansion in which case one can write 
\be 
L=\frac{2\tilde{K}}{\pi T}\left(1+\frac{\tilde{a}^{2}}{48}(5\log2-2) \right)\int\limits_{y_{t}}^{\infty}dy \frac{1}{y^{2}\sqrt{\left(y^{4}-y_{c}^{4}+\frac{\tilde{a}^{2}}{24}\La(y)\cosh^{2}\eta\right)}}+...
\ee
Also, in the limit $\eta$ becoming very large, $\La(y)$ reduces to $\La(y)=1-10\log2$, which is, in fact, independent of $y$. In this simplified scenario, the integral can be handled analytically and we have,
\be 
L=\frac{2\tilde{K}}{\pi T}\left(1+\frac{\tilde{a}^{2}}{48}(5\log2-2) \right)\frac{\sqrt{\pi}}{y_{t}^{3}}\frac{\G(3/4)}{\G(1/4)}.
\ee
It is now a straightforward exercise to compute the value of $\tilde{K}$ and hence, $y_{t}$ which maximize $L$ as
\bea \label{K}
\tilde{K}^{2}&=&2\cosh^{2}\eta+\frac{\tilde{a}^{2}\cosh^{2}\eta\left(10\log2-1 \right)}{12}\nn \\
y_{t}&=&y_{c}\left( 1+\frac{\tilde{a}^{2}\left(10\log2-1 \right)}{96}\cosh^{2}\eta\right).
\eea
Incorporating these values we arrive at the final expression for the screening length $L_{max}$ as,
\bea \label{Lmax}
L_{max}&=&\frac{1}{\sqrt{\pi}T}\frac{\G(3/4)}{\G(1/4)}\frac{2\sqrt{2}}{3^{3/4}}\frac{1}{\sqrt{\cosh \eta}}\left( 1-\frac{\tilde{a}^{2}}{16}(2.965735\cosh^{2}\eta-0.488578)\right)\nn \\
&=&\frac{1}{\sqrt{\pi}T}\frac{\G(3/4)}{\G(1/4)}\frac{2\sqrt{2}}{3^{3/4}}(1-v^{2})^{1/4}\left( 1-\frac{\tilde{a}^{2}}{16}\left(\frac{2.965735}{(1-v^{2})}-0.488578\right)\right)
\eea
The proportional change brought by anisotropy is,
\be 
\frac{\Delta L_{max}}{L_{max\big{|}_{\tilde{a}=0}}}=-\frac{\tilde{a}^{2}}{16}(2.965735\cosh^{2}\eta-0.488578).
\ee
Having deduced the analytical expression for $L_{max}$, a few comments are in order here. First, as expected, by setting $\tilde{a}=0$ here one recovers the usual screening length in an isotropic plasma \cite{Raj,Liu2}.  Second, it is obvious that the correction factor is always negative so that $L_{max}$ decreases in the presence of anisotropy. Third, when $\tilde{a}$ increases, the fall in $L_{max}$ is greater. Again, keeping $\tilde{a}$ fixed, if $\eta$ increases, $L_{max}$ falls. These conclusions drawn from the analytic expression (\ref{Lmax}) are in agreement with all our numerical results in the $L(\tilde{K})$-$\tilde{K}$ plots discussed earlier. Observe that the correction in the screening length arising due to the presence of anisotropy depends on the rapidity parameter as well. One also finds that $L_{max}$ depends inversely upon the temperature  and 
scales with velocity as $(1-v^{2})^{1/4}$. The velocity-scaling obtained here is in agreement with that found in \cite{Cher3}, where, of course, arbitrary orientation of the dipole with respect to its velocity was allowed and the analysis was not restricted only to weak anisotropy. One infers from (\ref{K}) that the value of $\tilde{K}$ which maximizes $L$ increases when we turn on the anisotropy parameter. This is also nicely exposed in the $L(\tilde{K})$-$\tilde{K}$ plot in Figs.\ref{12eta1} and \ref{12eta4} where the peaks gradually shift towards right as the anisotropy gets larger. With this we close our discussion of this configuration and move over to the next case.

\subsection{Motion in transverse plane, dipole along $x^{3}$} \label{13}
\begin{figure}[t]
\begin{center}
\subfigure[]{
\includegraphics[width=7.2cm,height=7.2cm, angle=-0]{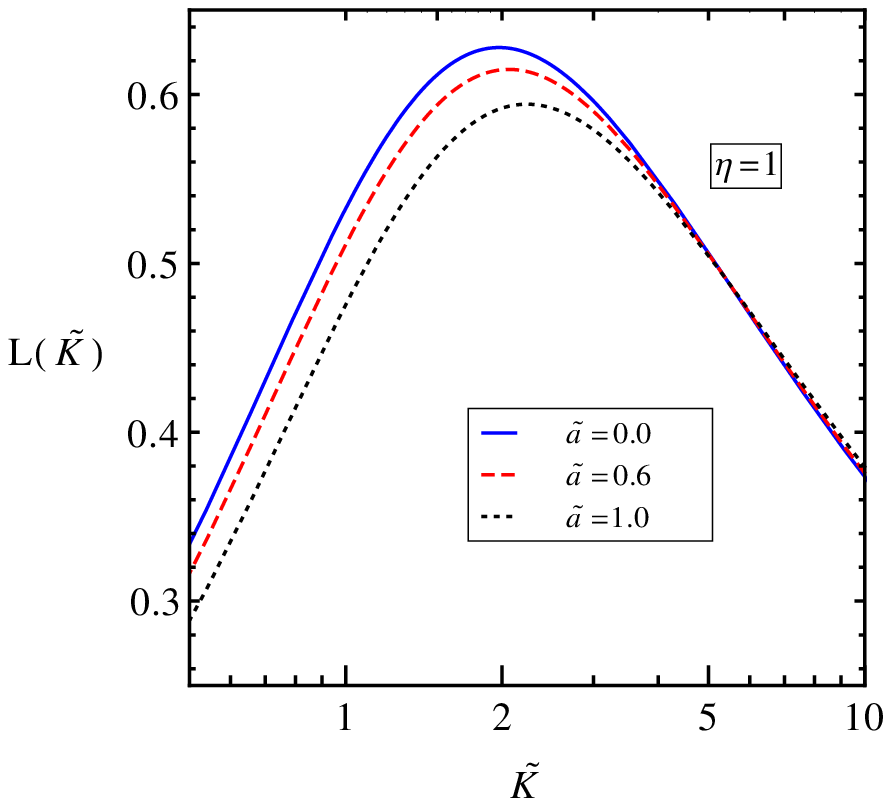}}
\hspace{5mm}
\subfigure[]{
\includegraphics[width=7.2cm,height=7.2cm, angle=-0]{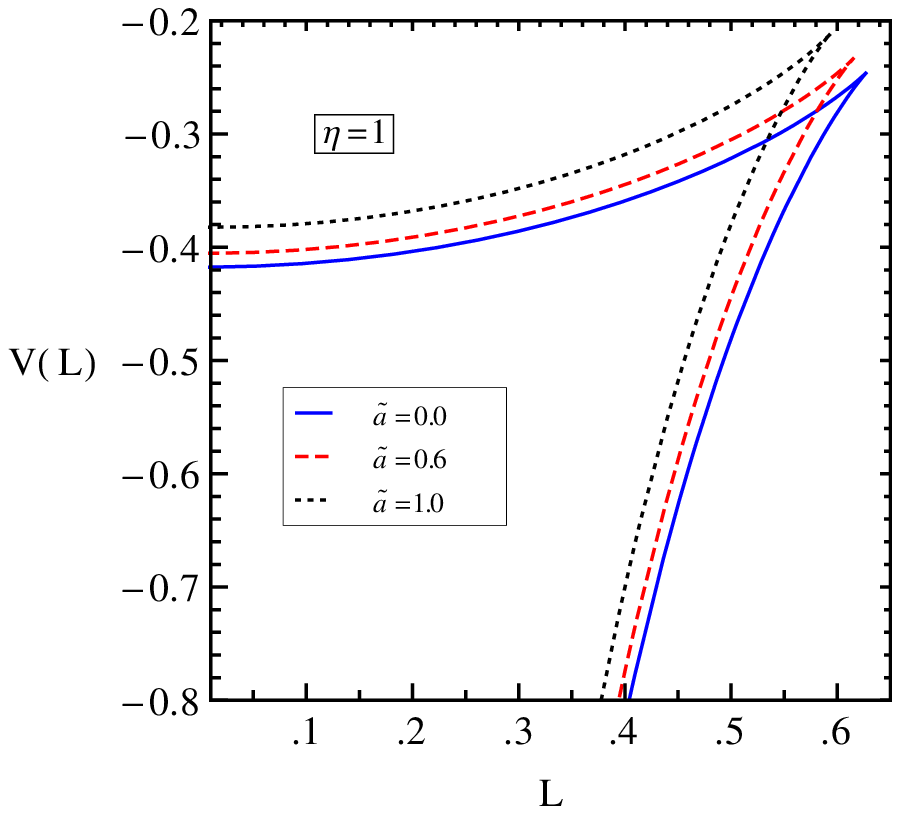}}
\caption{\label{13eta1} \small{(a) shows the plot of 
$L$ (normalized) as a function of $\tilde{K}$ with $\eta=1$ for
different values of $\tilde{a}$ when the velocity is in the transverse plane and dipole lies along anisotropic direction. (b) shows the plot
of properly normalized $V$ as a function of $L$ with
$\eta=1$ for the same set of values of $\tilde{a}$.}}
\end{center}
\end{figure}
\begin{figure}[t]
\begin{center}
\subfigure[]{
\includegraphics[width=7.1cm,height=7.1cm, angle=-0]{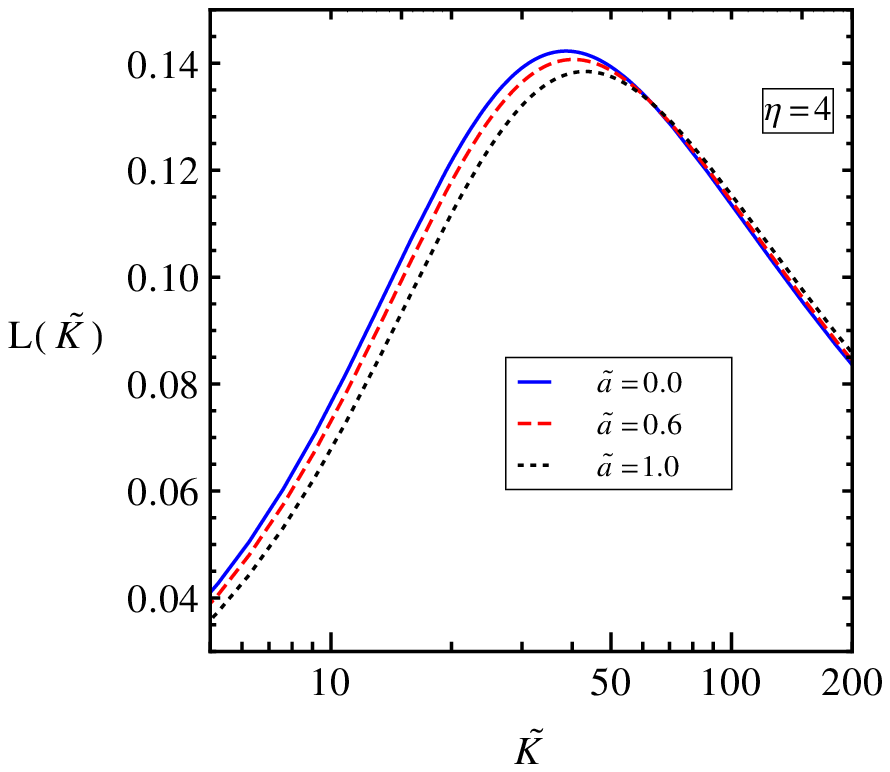}}
\hspace{5mm}
\subfigure[]{
\includegraphics[width=7.1cm,height=7.1cm, angle=-0]{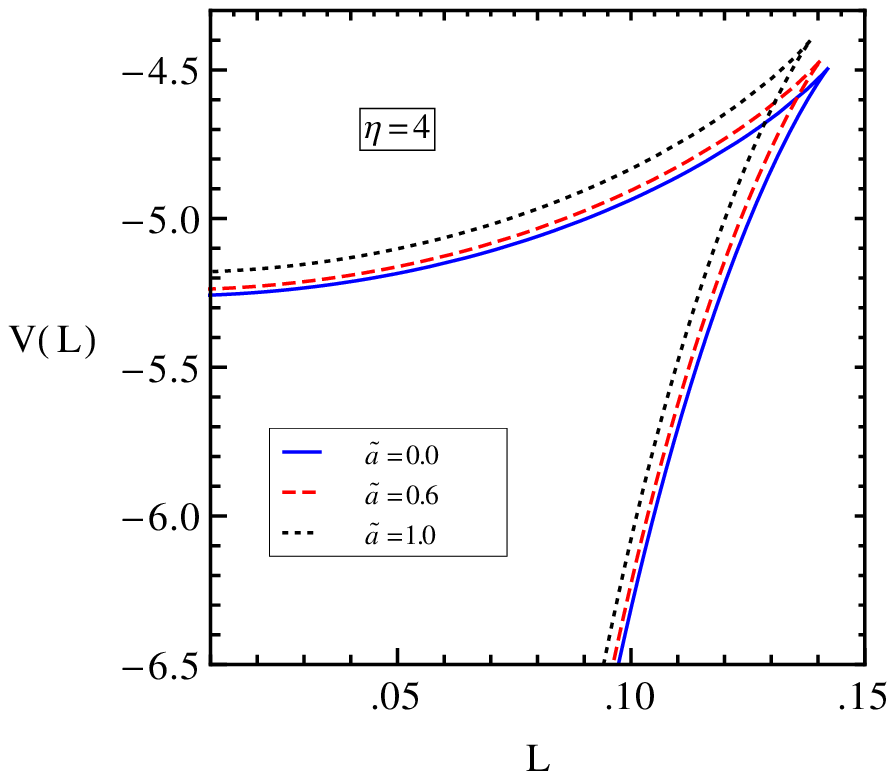}}
\caption{\label{13eta4} \small{(a) shows the plot of 
$L$ (normalized) as a function of $\tilde{K}$ with $\eta=4$ for
different values of $\tilde{a}$ when the velocity is in the transverse plane and dipole lies along anisotropic direction. (b) shows the plot
of properly normalized $V$ as a function of $L$ with
$\eta=4$ for the same set of values of $\tilde{a}$.}}
\end{center}
\end{figure}
In this case the dipole lies along the anisotropic direction $x^{3}$ and moves in the transverse plane with a velocity $v$.  Without any loss of generality, we can take the direction of motion to be along $x^{1}$. The calculation in this case proceeds in pretty much the same way. So we shall be brief in this section, pointing out only the differences that crops up in the calculations as we go along. Firstly, note that the choice of the static gauge is slightly altered. Now we take $\tau=t, \si=x^{3}, r=r(\si)$ with $x^{1,2}$ being independent of $\tau$ or $\si$. Enforcing this choice of gauge in the Nambu-Goto action (\ref{NG}) results in the following form of the action
\be 
S=\frac{\z{T}r_{h}}{2 \pi \a '}\int\limits_{-l/2}^{+l/2} d\tilde{\si}\sqrt{\left(A\left( G_{33}+G_{rr}y'^{2}\right) \right)}.
\ee
As before, the absence of any explicit $\tilde{\si}$-dependence leads to the following conserved quantity,
\be 
K=\frac{AG_{33}}{\sqrt{A\left( G_{33}+G_{rr}y'^{2}\right)}}
\ee
and the scaled \qq separation assumes the form,
\be 
l=2K\int\limits_{y_{t}}^{\infty}dy\sqrt{\frac{G_{rr}}{G_{33}}}\frac{1}{\sqrt{AG_{33}-K^{2}}}.
\ee
\begin{figure}[t]
\begin{center}
\subfigure[]{
\includegraphics[width=7.1cm,height=7.1cm, angle=-0]{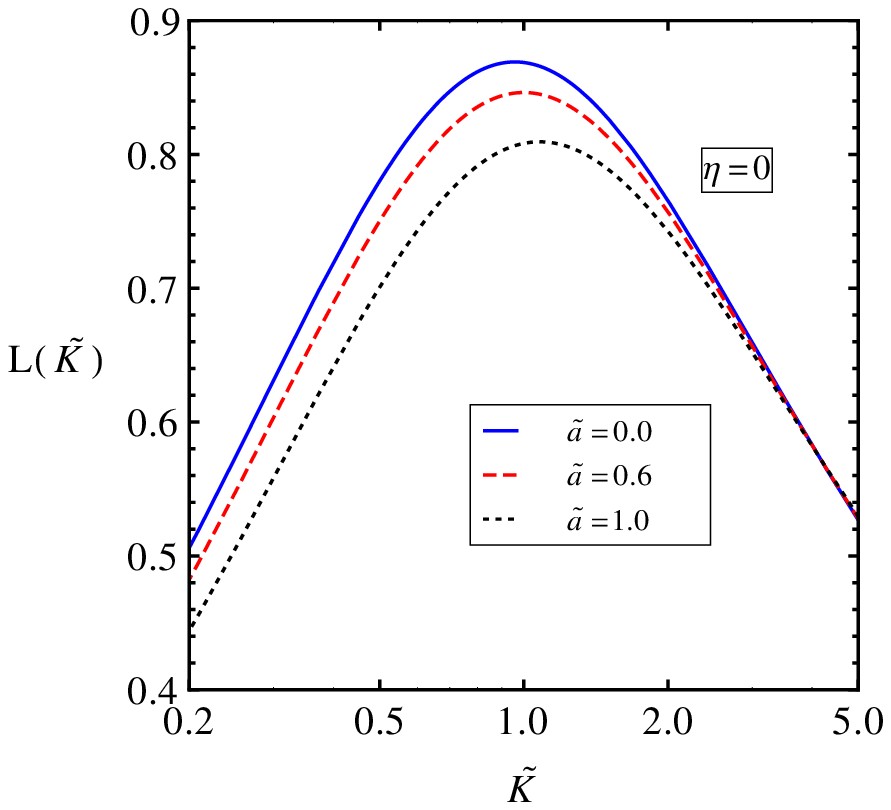}}
\hspace{5mm}
\subfigure[]{
\includegraphics[width=7.4cm,height=7.4cm, angle=-0]{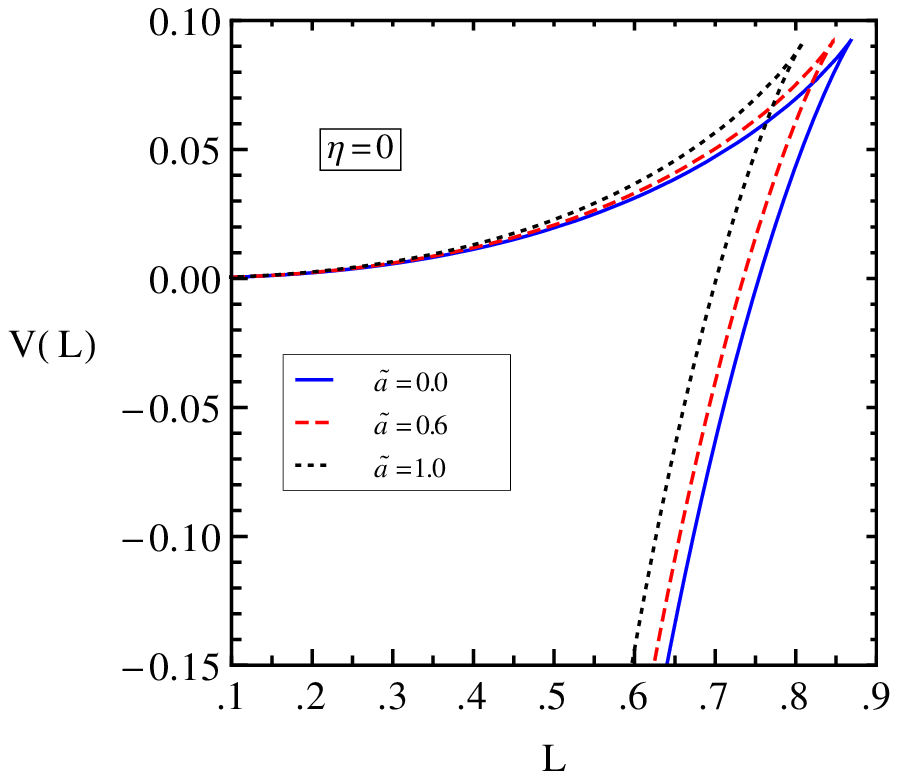}}
\caption{\label{13eta0} \small{(a) shows the plot of 
$L$ (normalized) as a function of $\tilde{K}$ with $\eta=0$ for
different values of $\tilde{a}$ when the dipole lies along anisotropic direction. (b) shows the plot
of properly normalized $V$ as a function of $L$ with
$\eta=0$ for the same set of values of $\tilde{a}$.}}
\end{center}
\end{figure}
Plugging in the explicit expressions of the metric components, we finally obtain the \qq separation as
\be  \label{L13}
L=\frac{2\tilde{K}}{\pi T}\left(1+\frac{\tilde{a}^{2}(5\log2-2)}{48} \right)\int\limits_{y_{t}}^{\infty}dy\frac{\z{H}^{-1}}{\sqrt{\left( y^{4}-1+\frac{\tilde{a}^{2}}{24}\Si\right)\left(y^{4}-(1-\frac{\tilde{a}^{2}}{24}\La)\cosh^{2}\eta-\tilde{K}^{2}\z{H}^{-1} \right)}}
\ee
where we have suppressed the explicit $y$-dependence of $\Si(y),\La(y)$ and $\z{H}(y)$. The turning point $y_{t}$ is found out by demanding the vanishing of the second term in the denominator at $y_{t}$, i.e., from
\be 
y_{t}^{4}-\cosh^{2}\eta+\frac{\tilde{a}^{2}}{24}\La(y_{t})\cosh^{2}\eta-\tilde{K}^{2}\z{H}(y_{t})^{-1}=0.
\ee
The action is given by
\be 
S=\frac{\z{T}r_{h}}{\pi \a '}\int\limits_{y_{t}}^{\infty}dyA\sqrt{\frac{G_{33}G_{rr}}{AG_{33}-K^{2}}}
\ee
which can be rewritten as
\bea \label{S13}
S&=&\frac{\z{T}r_{h}}{\pi \a '}\int\limits_{y_{t}}^{\infty}dy\frac{y^{4}-\cosh^{2}\eta+\frac{\tilde{a}^{2}}{24}\La\cosh^{2}\eta}{\sqrt{\left(y^{4}-1+\frac{\tilde{a}^{2}}{24}\Si \right)\left(y^{4}-(1-\frac{\tilde{a}^{2}}{24}\La)\cosh^{2}\eta-\tilde{K}^{2}\z{H}^{-1} \right)}} \nn \\
&\equiv&\frac{\z{T}r_{h}}{\pi \a '}\int\limits_{y_{t}}^{\infty}dy\z{S}^{ani}.
\eea
To evaluate the \qq potential one also needs to subtract the self-energy term $S_{0}$. It is easy to convince oneself that in this case the expression for $S_{0}$ as given in (\ref{S012}) remains unaltered and the \qq potential will be given by (\ref{Vfin}) with $\z{S}^{ani}$ now taken to be as in (\ref{S13}). We have given the $L(\tilde{K})$-$K$ and the $V(L)$-$L$ plots in Figs.\ref{13eta1} and \ref{13eta4} for $\eta=1$ and $\eta=4$ respectively. Fig.\ref{13eta0} shows the static \qq separation and the static \qq potential. We observe that in all the cases the general pattern of the plots (like the rightwards shift of the peak in the $L(\tilde{K})$ curves, attenuation of  $L_{max}$ and rise in the $V(L)$ plots with increasing $\tilde{a}$) mimic those obtained earlier in Sec.\ref{12} and hence does not merit a separate discussion.

\subsection{Motion along $x^{3}$, dipole in transverse plane}\label{31}
Third in our list is the case where the dipole is aligned in the transverse plane  and it has a velocity along the anisotropic direction. For the sake of simplicity we have taken the dipole to lie along $x^{1}$. While we shall proceed along the same line as in the previous cases, this time the calculations will be a little different since we now need to give a boost along the anisotropic direction, $x^{3}$.  First of all, we go to the rest-frame $(t',x^{3\prime})$ of the \qq pair by inflicting the boost
\be 
\begin{aligned}
& dt=\cosh \eta dt'-\sinh \eta dx^{3\prime},\\
& dx^{3}=-\sinh \eta dt'+\cosh \eta dx^{3\prime}.
\end{aligned}
\ee
The Wilson  loop so  formed spans the $t'$ and $x^{1}$ directions. In terms of the boosted coordinates the metric (\ref{MT}) can be rewritten as
\bea \label{MT31boosted}
ds^{2}\!\!\!\!&=&\!\!\!\!-\tilde{A}(r)dt^{2}-2\tilde{B}(r)dtdx^{3}+\tilde{C}(r)(dx^{3})^{2}+r^{2}\left((dx^{1})^{2}+(dx^{2})^{2} +\frac{dr^{2}}{r^{4}\z{F}}\right)+e^{\half \phi}d\Om_{5}^{2} \nn\\
&=&\tilde{G}_{\mu \nu}dx^{\mu}dx^{\nu}
\eea
where
\bea
\tilde{A}(y)&=&\left(\frac{r_{h}}{y}\right)^{2}\left[ y^{4}-\cosh^{2}\eta+\frac{\tilde{a}^{2}}{24}\La\cosh^{2}\eta +y^{4}\sinh^{2}\eta(1-\z{H})\right],\\
\tilde{B}(y)&=&\left(\frac{r_{h}}{y}\right)^{2}\sinh\eta \cosh \eta\left[1-\frac{\tilde{a}^{2}}{24}\La+y^{4}(\z{H}-1)\right],\\
\tilde{C}(y)&=&\left(\frac{r_{h}}{y}\right)^{2}\left[ y^{4}+\sinh^{2}\eta-\frac{\tilde{a}^{2}}{24}\La\sinh^{2}\eta +y^{4}\cosh^{2}\eta(\z{H}-1)\right].
\eea
\begin{figure}[t]
\begin{center}
\subfigure[]{
\includegraphics[width=7.3cm,height=7.3cm, angle=-0]{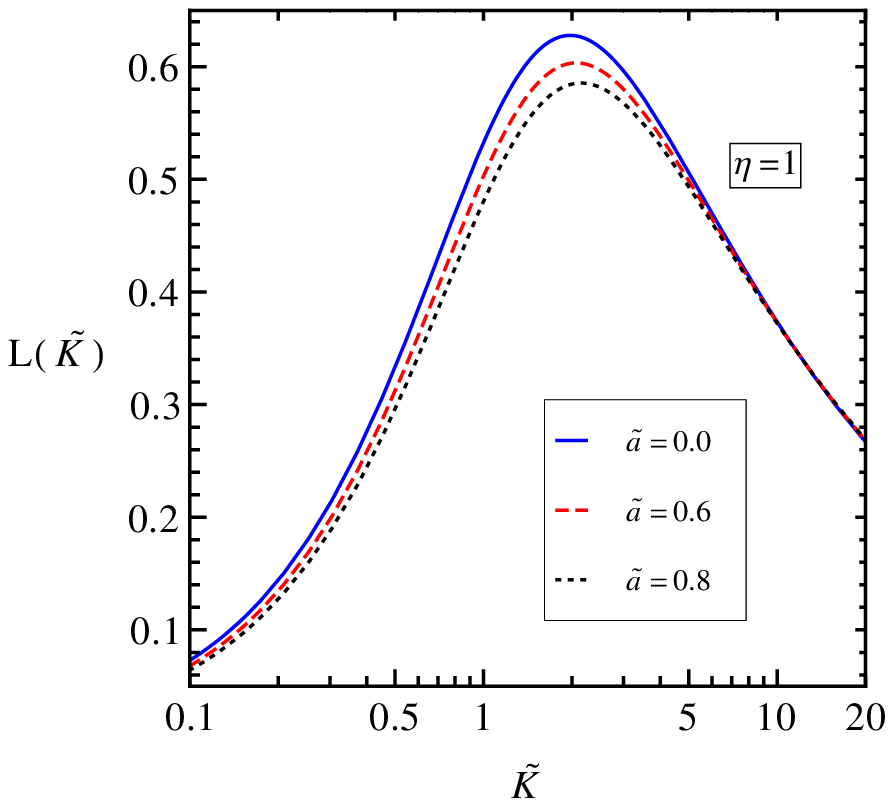}}
\hspace{5mm}
\subfigure[]{
\includegraphics[width=7.4cm,height=7.4cm, angle=-0]{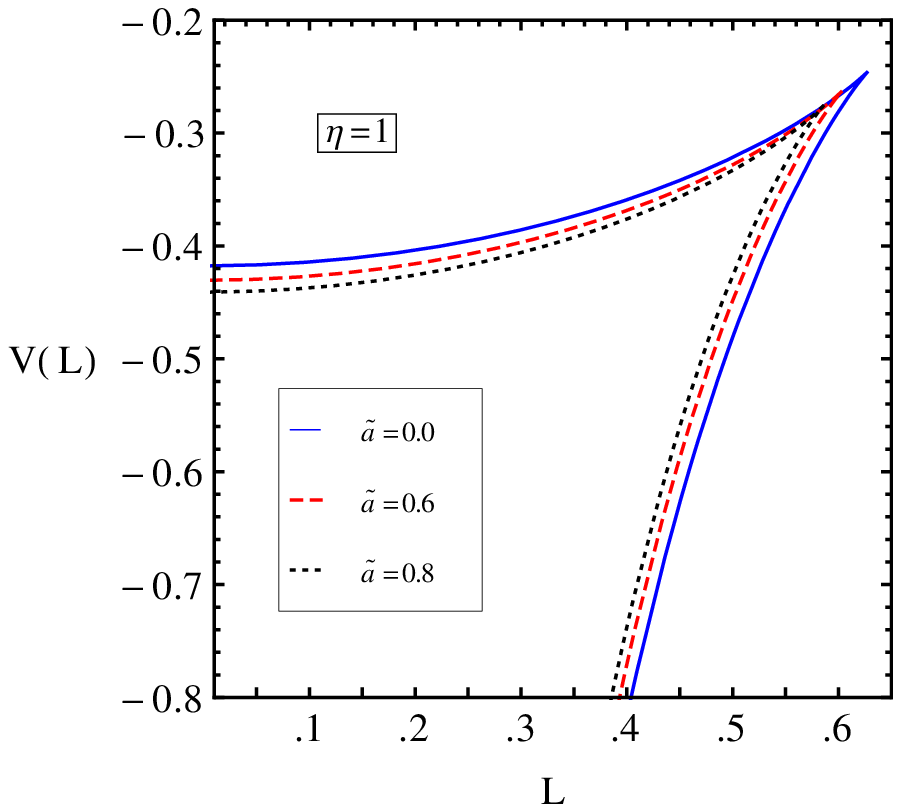}}
\caption{\label{31eta1} \small{(a) shows the plot of 
$L$ (normalized) as a function of $\tilde{K}$ with $\eta=1$ for
different values of $\tilde{a}$ when the velocity is along anisotropic direction and dipole lies in the transverse plane. (b) shows the plot
of properly normalized $V$ as a function of $L$ with
$\eta=1$ for the same set of values of $\tilde{a}$.}}
\end{center}
\end{figure}
To evaluate the Nambu-Goto string world-sheet action we employ the following choice of gauge: $\tau=t, \si=x^{1}, r=r(\si)$ with $x^{2,3}$ having no $\tau$- or $\si$-dependence. The action (\ref{NG}) can now be written as
\be \label{NG31} 
S=\frac{\z{T}r_{h}}{2 \pi \a '}\int\limits_{-l/2}^{+l/2} d\tilde{\si}\sqrt{\tilde{A}\left( \tilde{G}_{11}+ \tilde{G}_{rr}y'^{2}\right)}
\ee
Again the absence of any explicit $\si$-dependence furnishes the conserved quantity,
\be 
K=\frac{\tilde{A} \tilde{G}_{11}}{\sqrt{\tilde{A}\left(  \tilde{G}_{11}+ \tilde{G}_{rr}y'^{2}\right)}}.
\ee
Proceeding in the same way as before we get the scaled \qq separation,
\be 
l=2K\int\limits_{y_{t}}^{\infty}dy\sqrt{\frac{ \tilde{G}_{rr}}{ \tilde{G}_{11}}}\frac{1}{\sqrt{\tilde{A} \tilde{G}_{11}-K^{2}}}
\ee
from which one can read off the actual \qq separation
\bea  \label{L31}
L&=&\frac{2\tilde{K}}{\pi T}\left(1+\frac{\tilde{a}^{2}(5\log2-2)}{48} \right)\int\limits_{y_{t}}^{\infty}dy\frac{1}{\sqrt{\left( y^{4}-1+\frac{\tilde{a}^{2}}{24}\Si\right)}} \times \nn \\
&~&\hspace{1.5cm}\frac{1}{\sqrt{y^{4}-y_{c}^{4}+\frac{\tilde{a}^{2}}{24}\La\cosh^{2}\eta+y^{4}\sinh^{2}\eta (1-\z{H})}}.
\eea
The turning point $y_{t}$ is found from the solution of
\be \label{yt31}
y_{t}^{4}-y_{c}^{4}+\frac{\tilde{a}^{2}}{24}\La(y_{t})\cosh^{2}\eta+y_{t}^{4}\sinh^{2}\eta (1-\z{H}(y_{t}))=0.
\ee
\begin{figure}[t]
\begin{center}
\subfigure[]{
\includegraphics[width=7.4cm,height=7.4cm, angle=-0]{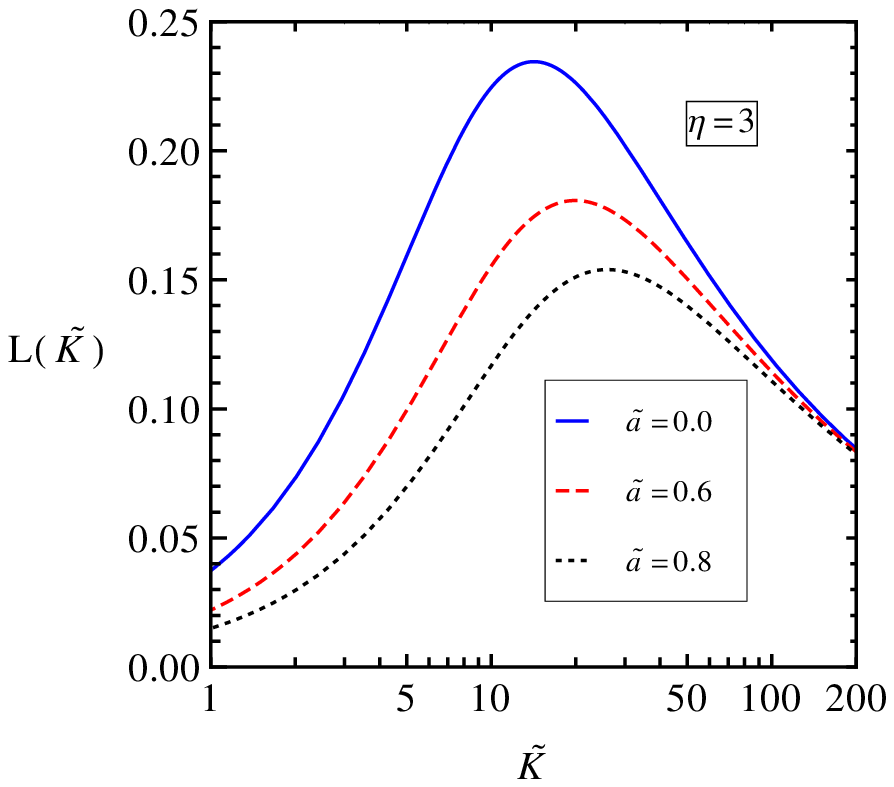}}
\hspace{5mm}
\subfigure[]{
\includegraphics[width=7.5cm,height=7.4cm, angle=-0]{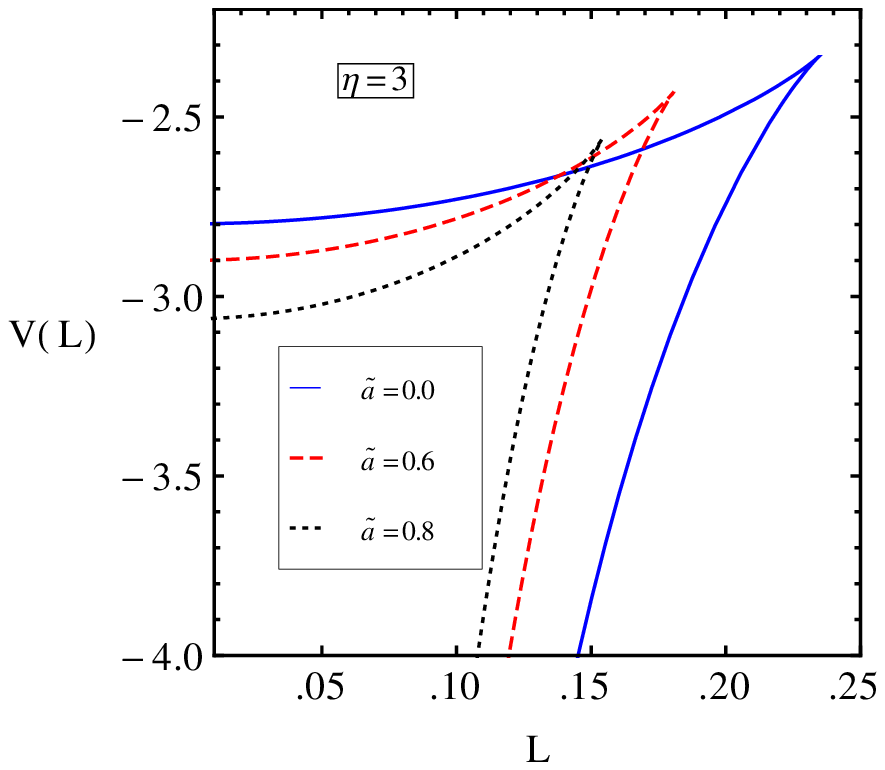}}
\caption{\label{31eta3} \small{(a) shows the plot of 
$L$ (normalized) as a function of $\tilde{K}$ with $\eta=3$ for
different values of $\tilde{a}$ when the velocity is along anisotropic direction and dipole lies in the transverse plane. (b) shows the plot
of properly normalized $V$ as a function of $L$ with
$\eta=3$ for the same set of values of $\tilde{a}$.}}
\end{center}
\end{figure}
The factor $(1-\z{H})$ goes as $\frac{\tilde{a}^{2}}{4}\log \left(1+\frac{1}{y^{2}}\right)$ up to $\z{O}(\tilde{a}^{2})$ and for large $y$ its contribution to the second factor in the denominator is $ \frac{\tilde{a}^{2}}{4}y^{2}\sinh^{2}\eta$. This is greater than the other anisotropic term by $\z{O}(y^{2})$ for large $y$. Hence, unlike in the previous cases, this time we do not expect the turning point $y_{t}$ to appear in the form of a correction to the isotropic value $y_{c}$ since the presence of this $\z{O}(y^{2})$ term renders the applicability of perturbative methods to solve the above equation futile. Thus, one has to depend solely upon numerical techniques to solve (\ref{yt31}) in order to extract $y_{t}$. In fact, numerical evaluation shows $y_{t}$ to be markedly different from $y_{c}$, particularly for low values of $\tilde{K}$. Once we have obtained $y_{t}$, we use it in (\ref{L31}) to 
numerically study the \qq separation.
The string world-sheet action is
\be 
S=\frac{\z{T}r_{h}}{\pi \a '}\int\limits_{y_{t}}^{\infty}dy\tilde{A}\sqrt{\frac{ \tilde{G}_{11} \tilde{G}_{rr}}{\tilde{A} \tilde{G}_{11}-K^{2}}}
\ee
which, written explicitly, assumes the following form,
\bea \label{S31}
S&=&\frac{\z{T}r_{h}}{\pi \a '}\int\limits_{y_{t}}^{\infty}dy\frac{y^{4}-\cosh^{2}\eta+\frac{\tilde{a}^{2}}{24}\La\cosh^{2}\eta+y^{4}\sinh^{2}\eta (1-\z{H})}{\sqrt{\left(y^{4}-1+\frac{\tilde{a}^{2}}{24}\Si \right)\left(y^{4}-y_{c}^{4}+\frac{\tilde{a}^{2}}{24}\La\cosh^{2}\eta+y^{4}\sinh^{2}\eta (1-\z{H})\right)}} \nn \\
&\equiv&\frac{\z{T}r_{h}}{\pi \a '}\int\limits_{y_{t}}^{\infty}dy\z{S}^{ani}.
\eea
As in the preceding cases, this action is divergent which is cured by taking away the self-energy contribution $S_{0}$ of the \qq pair. To compute $S_{0}$ we consider an open string hanging down the radial direction in the following gauge, $\tau=t, \si=r, x^{3}=x^{3}(\si)$ and $x^{1},x^{2}$ are independent of $\tau,\si$. Repeating the same exercise as in Sec.\ref{12} one finds $S_{0}$ to be of the form
\bea \label{S031}
S_{0}\!\!\!\!&=&\!\!\!\!\frac{\z{T}}{\pi \a '}\int\limits_{r_{h}}^{\infty} dr \sqrt{\tilde{A} \tilde{G}_{rr}}\sqrt{\frac{\tilde{A}\tilde{C}+\tilde{B}^{2}}{\tilde{A}\tilde{C}+\tilde{B}^{2}-K_{0}^{2}}}\nn \\
\!\!\!\!&=&\!\!\!\! \frac{\z{T}r_{h}}{\pi \a '}\int\limits_{1}^{\infty} dy \frac{\sqrt{y^{4}-\cosh^{2}\eta+\frac{\tilde{a}^{2}}{24}\La\cosh^{2}\eta+y^{4}\sinh^{2}\eta (1-\z{H})}}{\sqrt{y^{4}-1-\tilde{K}_{0}^{2}+\frac{\tilde{a}^{2}}{24}\La\z{H}+(\z{H}-1)(y^{4}-1)}}\times \nn \\
&~&\hspace{4cm}\frac{\sqrt{y^{4}-1+\frac{\tilde{a}^{2}}{24}\La\z{H}+(\z{H}-1)(y^{4}-1)}}{\sqrt{ y^{4}-1+\frac{\tilde{a}^{2}}{24}\Si}}\nn \\
\!\!\!\!&\equiv&\!\!\!\!\frac{\z{T}r_{h}}{\pi \a '}\int\limits_{1}^{\infty} dy \z{S}^{ani}_{0}
\eea
where $K_{0}$ is the conserved quantity owing its origin to the absence of any explicit $x^{3}$-dependence in the action. The second terms each in the numerator and the denominator separately vanish at $y=1$ providing a potential turning point  $y_{t}=1$. The first term in the denominator can contribute another turning point $y_{t}>1$ but that possibility is ruled out by judiciously choosing the constant $\tilde{K}_{0}$ such that the zero of the first term in the numerator coincides with that of the first term in the denominator. We are now in a position to finally compute the \qq potential (\ref{Vfin}) with $\z{S}^{ani}$ provided in (\ref{S31}) and the corresponding self-energy term $\z{S}_{0}^{ani}$ in (\ref{S031}). Using the above information we have plotted the \qq separation and the \qq potential in Figs.\ref{31eta1} and \ref{31eta3}\footnote{Note that we have not given the static \qq separation and the static \qq potential in this case since these will be the same as in Sec.\ref{12}.}. While the gross 
features of the plots remain almost unaltered, observe that all the signatures of the presence of anisotropy are far more pronounced (particularly in the high rapidity regime)  than in either of the preceding cases. This has its roots in the presence of the $\z{O}(y^{2})$ term in the anisotropic contribution to the \qq separation and the \qq potential as mentioned earlier.  The heavy \qq potential for this configuration has also been found in \cite{Cher3}, using different values of the parameters and we find that our results tally with those presented in \cite{Cher3}. 
\begin{figure}[t]
\begin{center}
\subfigure[]{
\includegraphics[width=7.3cm,height=7.3cm, angle=-0]{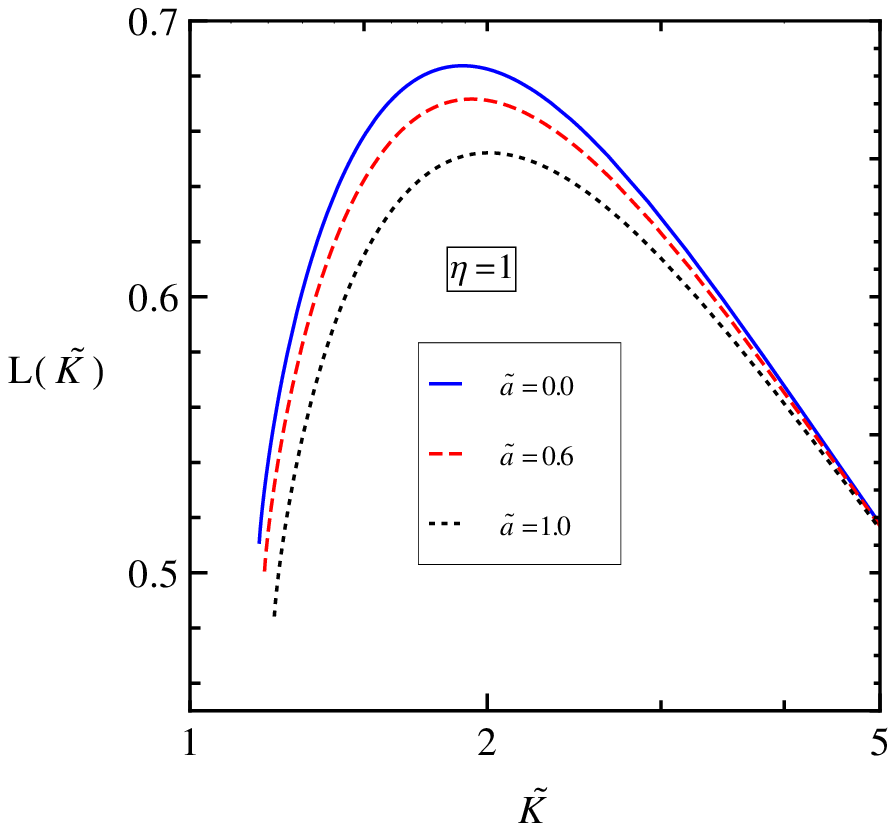}}
\hspace{5mm}
\subfigure[]{
\includegraphics[width=7.5cm,height=7.5cm, angle=-0]{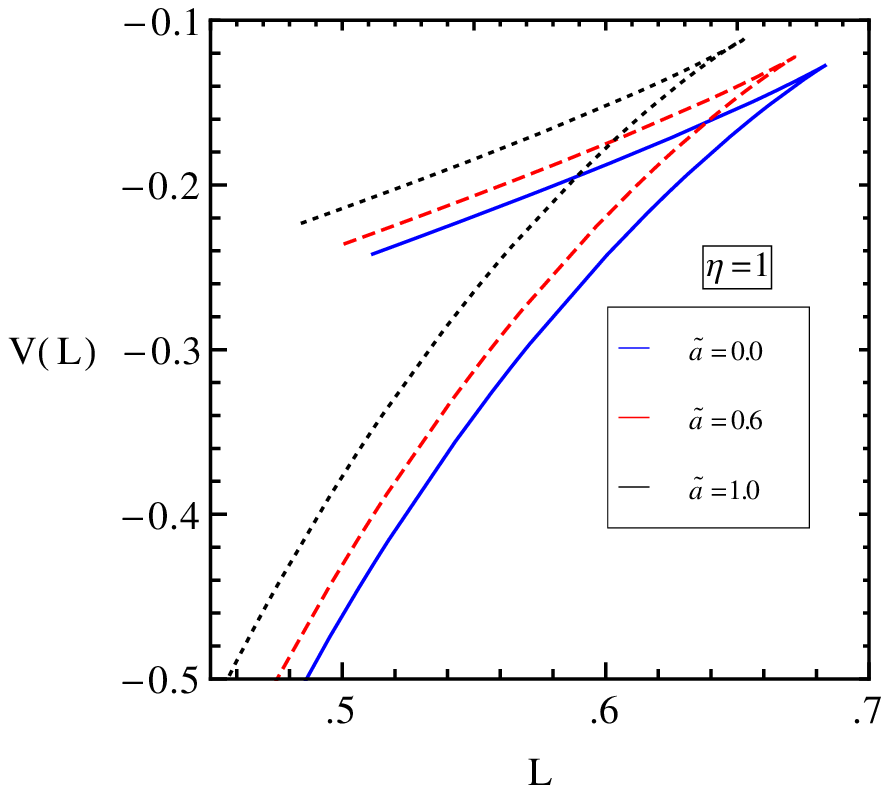}}
\caption{\label{11eta1} \small{(a) shows the plot of 
$L$ (normalized) as a function of $\tilde{K}$ with $\eta=1$ for
different values of $\tilde{a}$ when the dipole is parallel to its velocity and both lie in the transverse plane. (b) shows the plot
of properly normalized $V$ as a function of $L$ with
$\eta=1$ for the same set of values of $\tilde{a}$.}}
\end{center}
\end{figure}

\subsection{Motion in transverse plane, dipole parallel to direction of motion in the plane}\label{11}
We now come to the case where the dipole is aligned parallel to its direction of motion. This common direction can be in the transverse plane or along the anisotropic direction. We consider the former case in this subsection. For simplicity we shall take this common direction to be along $x^{1}$. Boosting to the rest-frame, choosing the static gauge, $\tau=t, \si=x^{1}, r=r(\si)$ and $x^{2}=x^{3}=\text{constant}$  leads us to the action,
\be 
S=\frac{\z{T}r_{h}}{2\pi \a'}\int\limits_{-l/2}^{l/2} d\tilde{\si}\sqrt{A\left(C+G_{rr}y'^{2} \right)+B^{2}}
\ee
which, in turn, supplies the constant of motion,
\be 
K=\frac{AC+B^{2}}{\sqrt{A\left(C+G_{rr}y'^{2} \right)+B^{2}}}.
\ee
Proceeding along the lines of the earlier cases, we compute,
\be \label{y'11}
y'=\frac{r_{0}^{2}}{\tilde{K}}\frac{\sqrt{\left( y^{4}-1+\frac{\tilde{a}^{2}}{24}\Si \right) \left( y^{4}-1+\frac{\tilde{a}^{2}}{24}\La\right)\left(y^{4}-1-\tilde{K}^{2}+\frac{\tilde{a}^{2}}{24}\La \right)}}{\sqrt{y^{4}-\cosh^{2}\eta+\frac{\tilde{a}^{2}}{24}\La\cosh^{2}\eta }}
\ee
from which we find the \qq separation to be
\bea \label{L11}
L&=&\frac{2\tilde{K}}{\pi T}\left(1+\frac{\tilde{a}^{2}(5\log2-2)}{48}     \right)\int\limits_{y_{t}}^{\infty}dy\frac{\sqrt{y^{4}-\cosh^{2}\eta+\frac{\tilde{a}^{2}}{24} \La\cosh^{2}\eta}}{\sqrt{\left( y^{4}-1+\frac{\tilde{a}^{2}}{24}\Si \right) \left( y^{4}-1+\frac{\tilde{a}^{2}}{24}\La\right)}}\times \nn \\
&~&\hspace{7.6cm}\frac{1}{\sqrt{y^{4}-1-\tilde{K}^{2}+\frac{\tilde{a}^{2}}{24}\La}}.
\eea
The turning point $y_{t}$ is obtained from (\ref{y'11}) which satisfies
\be 
y^{4}_{t}-1-\tilde{K}^{2}+\frac{\tilde{a}^{2}}{24}\La\left(y_{t}\right)=0.
\ee
At the same time, note that $y'$ now encounters a singularity at $y_{s}$, given by,
\be 
y_{s}^{4}-\cosh^{2}\eta+\frac{\tilde{a}^{2}}{24} \La(y_{s})\cosh^{2}\eta=0.
\ee
Further, it is evident that for $y<y_{s}$, the numerator in (\ref{L11}) becomes imaginary. So any potential turning point has to satisfy 
\be 
y_{t}^{4}-\cosh^{2}\eta+\frac{\tilde{a}^{2}}{24}\La(y_{t})\cosh^{2}\eta >0
\ee
which imposes a lower bound on $\tilde{K}$ that turns out to be\footnote{The existence of this lower bound is found in the isotropic case too as given in \cite{Liu}.},
\be  \label{Kbound}
\tilde{K}^{2}>\tilde{K}_{min}^{2}=\sinh^{2}\eta\left( 1-\frac{\tilde{a}^{2}}{24}\La\left(y=\sqrt{\cosh \eta}\right)\right).
\ee
Incidentally, note that this lower bound turns out to be the same as the constant $\tilde{K}_{0}$ that appeared in Sec.\ref{12}.
\begin{figure}[t]
\begin{center}
\subfigure[]{
\includegraphics[width=7.5cm,height=7.5cm, angle=-0]{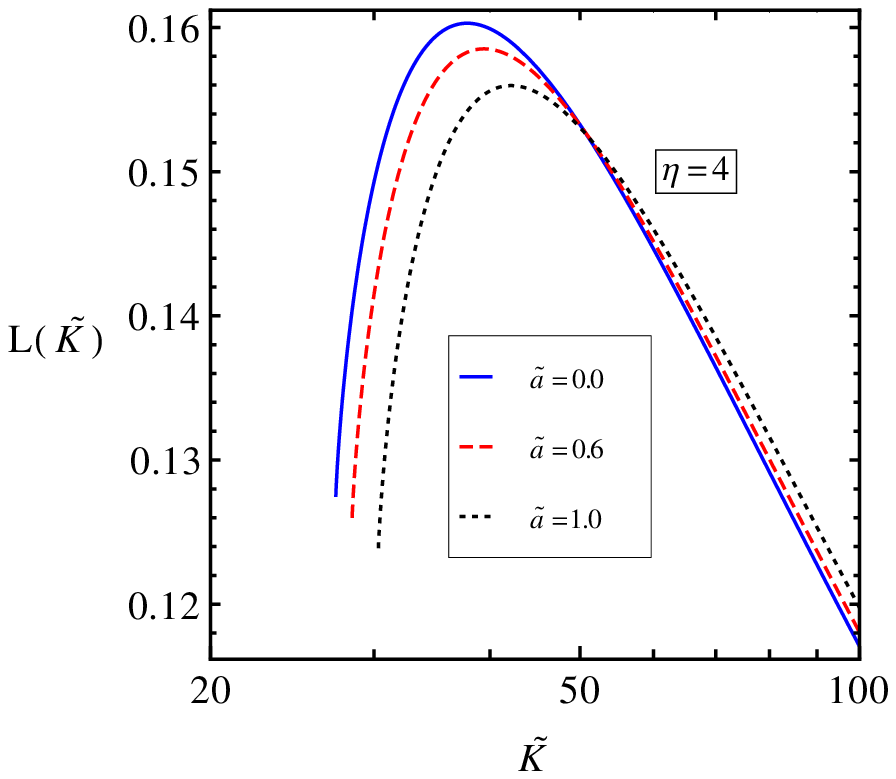}}
\hspace{5mm}
\subfigure[]{
\includegraphics[width=7.3cm,height=7.3cm, angle=-0]{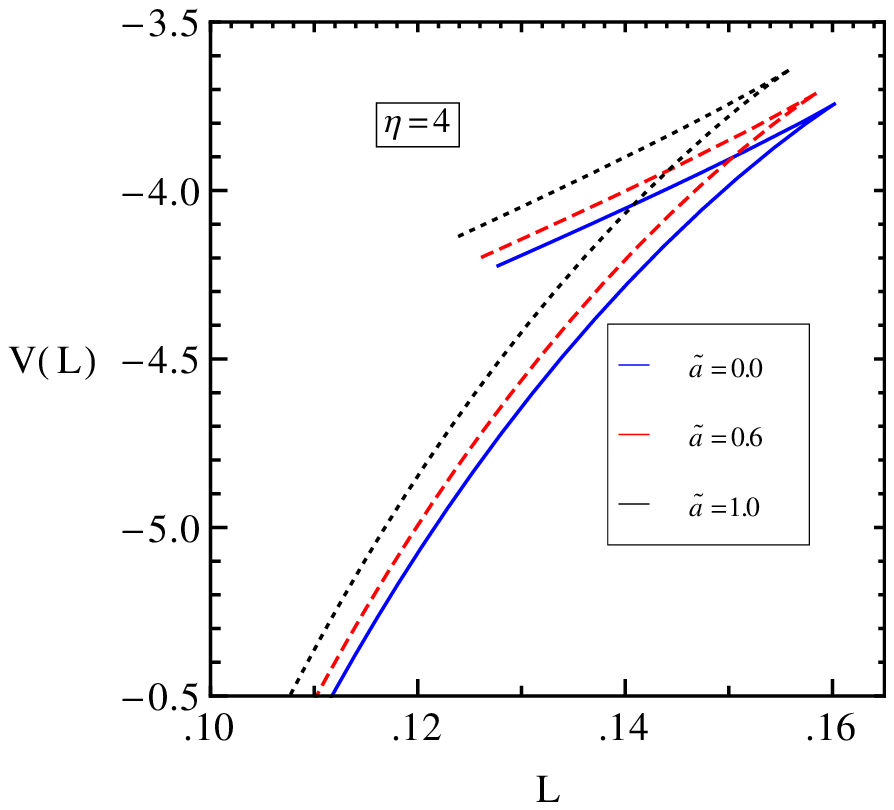}}
\caption{\label{11eta4} \small{(a) shows the plot of 
$L$ (normalized) as a function of $\tilde{K}$ with $\eta=4$ for
different values of $\tilde{a}$ when the dipole is parallel to its velocity and both lie in the transverse plane. (b) shows the plot
of properly normalized $V$ as a function of $L$ with
$\eta=4$ for the same set of values of $\tilde{a}$.}}
\end{center}
\end{figure}
Upon simplification the action boils down to,
\bea \label{S11}
S\!\!\!\!&=&\!\!\!\!\frac{\z{T}r_{h}}{\pi \a '}\int\limits_{y_{t}}^{\infty} dy \sqrt{AG_{rr}}\sqrt{\frac{AC+B^{2}}{AC+B^{2}-K^{2}}}\nn \\
\!\!\!\!&=&\!\!\!\! \frac{\z{T}r_{h}}{\pi \a '}\int\limits_{y_{t}}^{\infty} dy\frac{\sqrt{\left(y^{4}-\cosh^{2}\eta+\frac{\tilde{a}^{2}}{24}\La\cosh^{2}\eta\right)\left(y^{4}-1+\frac{\tilde{a}^{2}}{24}\La\right)}}{\sqrt{\left( y^{4}-1+\frac{\tilde{a}^{2}}{24}\Si\right) \left(y^{4}-1-\tilde{K}^{2} +\frac{\tilde{a}^{2}}{24}\La\right)}}\nn \\
\!\!\!\!&\equiv&\!\!\!\!\frac{\z{T}r_{h}}{\pi \a '}\int\limits_{y_{t}}^{\infty} dy \z{S}^{ani}
\eea
with $\tilde{K}$ respecting the inequality in (\ref{Kbound}). The self-energy contribution $S_{0}$ is also given by (\ref{S11}) but now with $\tilde{K}$ saturating the bound in (\ref{Kbound}) and the lower limit being $y=1$ so that $S_{0}$ becomes identical with that given in (\ref{S012}). We can now compute the \qq potential using (\ref{S012}) and (\ref{S11}) in (\ref{Vfin}) with (\ref{temp}). The \qq separation and the potential have been plotted in Figs.\ref{11eta1} and \ref{11eta4} for $\eta=1$ and $\eta=4$ respectively. The $L(\tilde{K})$-$\tilde{K}$ plots  show that curves for higher value of $\tilde{K}$ (after $L_{max}$ is attained) exhibits the same pattern as in the earlier cases but for lower values of $\tilde{K}$ there is an inaccessible region for $\tilde{K} \leq \tilde{K}_{min}$ for which there is no solution to the dipole separation. This is reflected in the $V(L)$-$L$ plot where the upper branch of the potential terminates abruptly at $L=L_{min}$ whereas the lower branch shows the usual 
behaviour . A closer scrutiny of the 
figures suggest that $\tilde{K}_{min}$ increases with increasing $\tilde{a}$ and concomitantly, $L_{min}$ decreases. However, this is manifested only in the unstable, high energy branch, which, in any case, is devoid of much physical significance, being energetically unfavorable.

\subsection{Motion and dipole, both along the anisotropic direction} \label{33}
\begin{figure}[t]
\begin{center}
\subfigure[]{
\includegraphics[width=7.3cm,height=7.3cm, angle=-0]{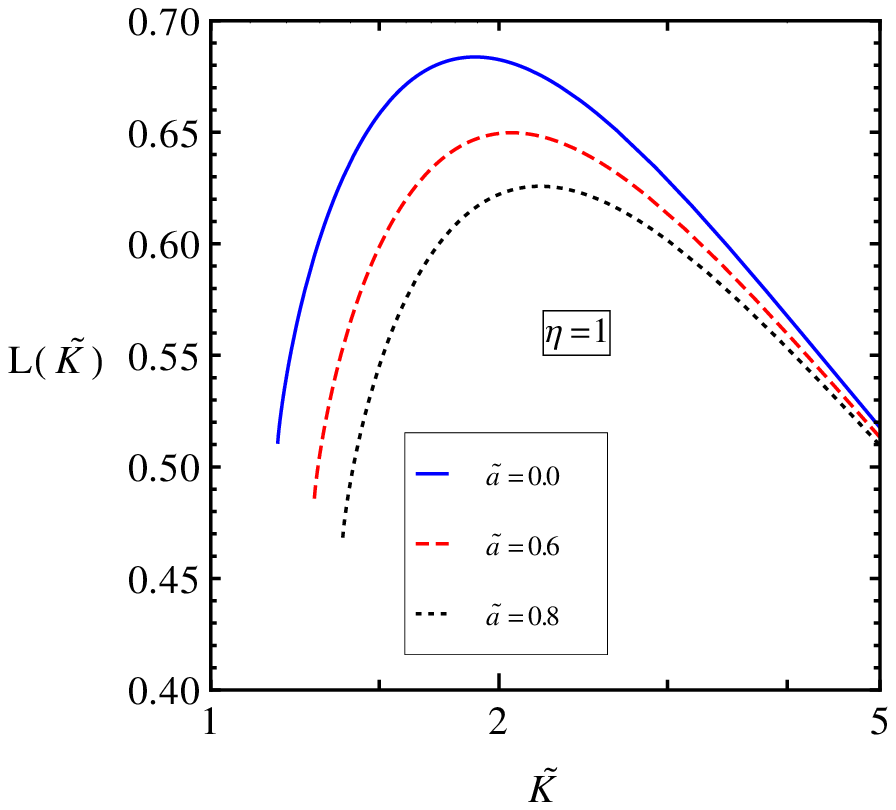}}
\hspace{5mm}
\subfigure[]{
\includegraphics[width=7.4cm,height=7.4cm, angle=-0]{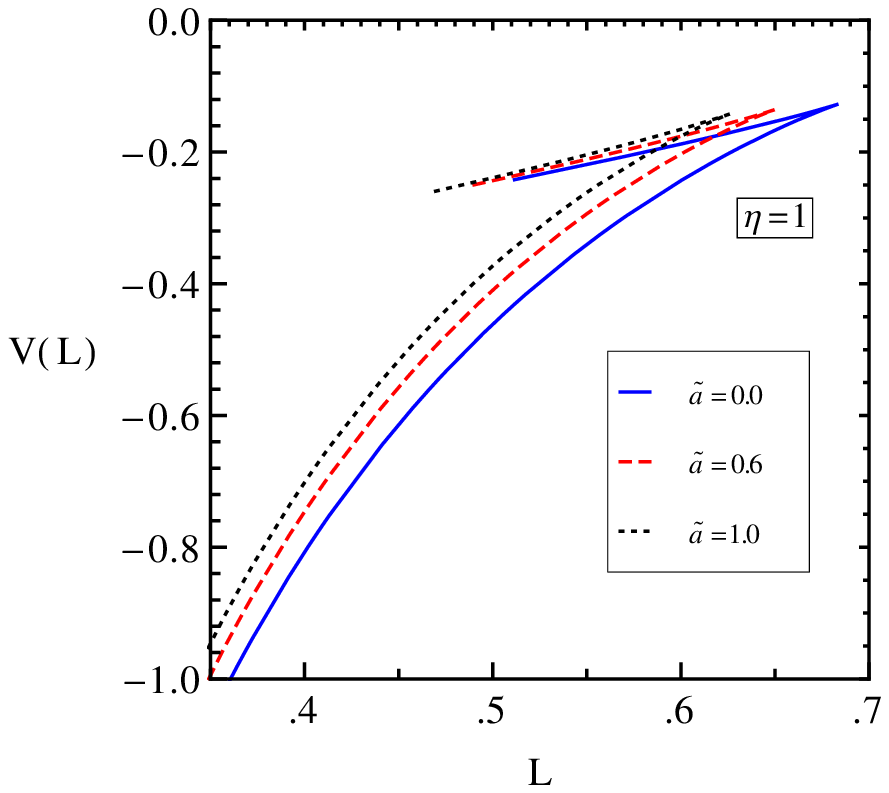}}
\caption{\label{33eta1} \small{(a) shows the plot of 
$L$ (normalized) as a function of $\tilde{K}$ with $\eta=1$ for
different values of $\tilde{a}$ when the dipole is parallel to its velocity and both lie along the anisotropic direction. (b) shows the plot
of properly normalized $V$ as a function of $L$ with
$\eta=1$ for the same set of values of $\tilde{a}$.}}
\end{center}
\end{figure}
Finally, we take up the case where the dipole is oriented along the anisotropic direction $x^{3}$ and it moves in the same direction. This time we shall make use of the metric (\ref{MT31boosted}) as obtained in Sec.\ref{31} and use the gauge choice of Sec.\ref{13}. All the calculations proceed in identically the same fashion as in Sec.\ref{11} and we end up with the \qq separation
\bea \label{L33}
L&\!\!\!\!\!\!\!\!=\!\!\!\!\!\!\!\!&\frac{2\tilde{K}}{\pi T}\left(1+\frac{\tilde{a}^{2}(5\log2-2)}{48}     \right)\int\limits_{y_{t}}^{\infty}dy\frac{\sqrt{y^{4}-\cosh^{2}\eta+\frac{\tilde{a}^{2}}{24}\La\cosh^{2}\eta +y^{4}\sinh^{2}\eta(1-\z{H})}}{\sqrt{y^{4}-1-\tilde{K}^{2}+\frac{\tilde{a}^{2}}{24}\La\z{H}+(\z{H}-1)(y^{4}-1)}}\times \nn \\
&~&\hspace{4.cm}\frac{1}{\sqrt{\left( y^{4}-1+\frac{\tilde{a}^{2}}{24}\Si \right) \left( y^{4}-1+\frac{\tilde{a}^{2}}{24}\La\z{H}+(\z{H}-1)(y^{4}-1)\right)}}.
\eea
The first term in the denominator provides the turning point $y_{t}$ which, in turn, is constrained by the condition that the numerator must be real. This results in a lower cut-off on the value of $\tilde{K}$. We can read off this lower bound by demanding that the zeros of the numerator and the first factor in the denominator occur at the same value of $y$. As was the case in Sec.\ref{31} due to the presence of the $y^{4}\sinh^{2}\eta(1-\z{H})$ term here we do not expect the effect of anisotropy to be small enough so as to employ perturbative methods. Hence, we have evaluated the lower limit $\tilde{K}_{min}$ and $y_{t}$ completely numerically.
Finally, the action becomes,
\bea \label{S33}
S\!\!\!\!&=&\!\!\!\! \frac{\z{T}r_{h}}{\pi \a '}\int\limits_{y_{t}}^{\infty} dy \frac{\sqrt{y^{4}-\cosh^{2}\eta+\frac{\tilde{a}^{2}}{24}\La\cosh^{2}\eta+y^{4}\sinh^{2}\eta (1-\z{H})}}{\sqrt{y^{4}-1-\tilde{K}^{2}+\frac{\tilde{a}^{2}}{24}\La\z{H}+(\z{H}-1)(y^{4}-1)}}\times \nn \\
&~&\hspace{4cm}\frac{\sqrt{y^{4}-1+\frac{\tilde{a}^{2}}{24}\La\z{H}+(\z{H}-1)(y^{4}-1)}}{\sqrt{ y^{4}-1+\frac{\tilde{a}^{2}}{24}\Si}}\nn \\
\!\!\!\!&\equiv&\!\!\!\!\frac{\z{T}r_{h}}{\pi \a '}\int\limits_{y_{t}}^{\infty} dy \z{S}^{ani}.
\eea
\begin{figure}[t]
\begin{center}
\subfigure[]{
\includegraphics[width=7.3cm,height=7.3cm, angle=-0]{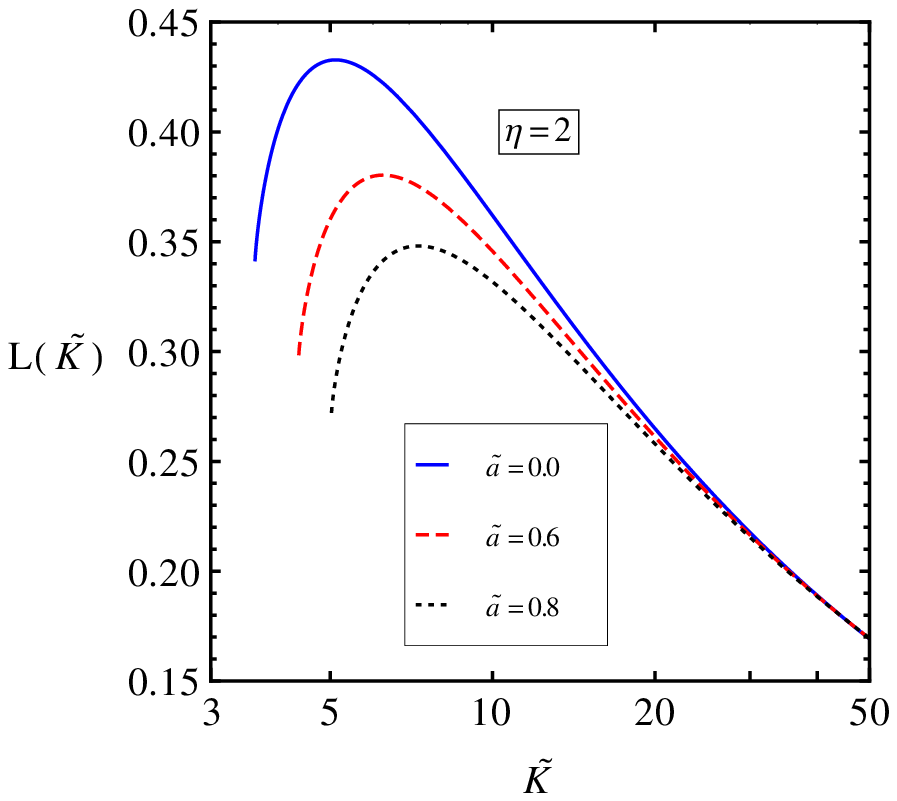}}
\hspace{5mm}
\subfigure[]{
\includegraphics[width=7.5cm,height=7.5cm, angle=-0]{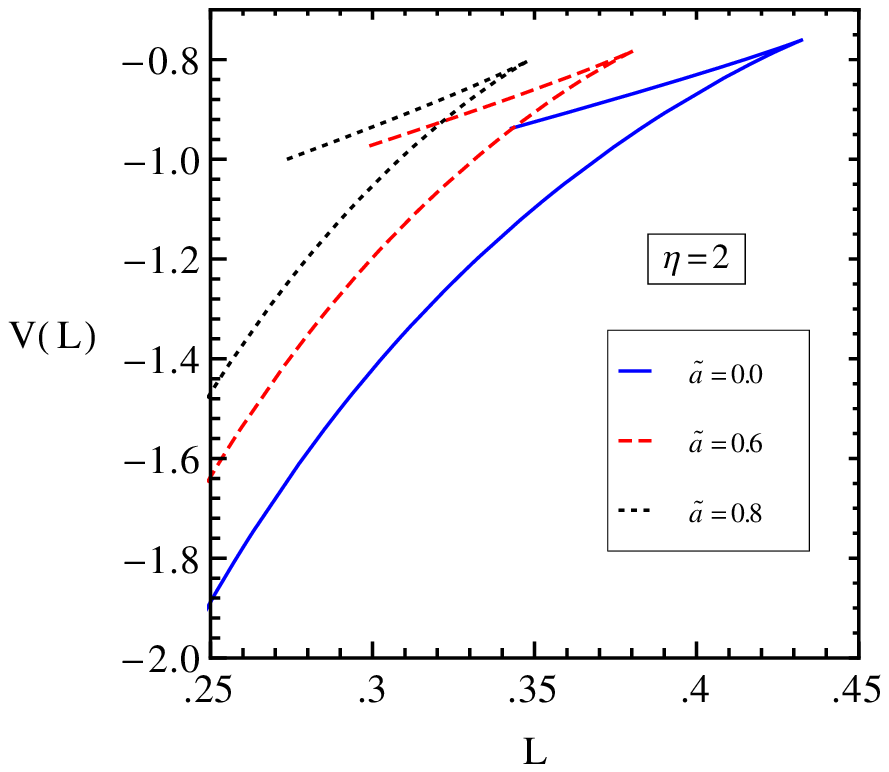}}
\caption{\label{33eta2} \small{(a) shows the plot of 
$L$ (normalized) as a function of $\tilde{K}$ with $\eta=2$ for
different values of $\tilde{a}$ when the dipole is parallel to its velocity and both lie along the anisotropic direction. (b) shows the plot
of properly normalized $V$ as a function of $L$ with
$\eta=2$ for the same set of values of $\tilde{a}$.}}
\end{center}
\end{figure}
In similar fashion, one finds the self-energy contribution $S_{0}$ to be the same as in (\ref{S031}) (and, in fact, (\ref{S33}) with $\tilde{K}$ replaced by its minimum value $\tilde{K}_{min}$ and lower limit at $y=1$). Equipped with this much information we can now obtain the plots for the dipole separation and the potential which are given in Fig.\ref{33eta1} and \ref{33eta2} for $\eta=1$ and $\eta=2$ (corresponding to $v=0.964$) respectively. The plots are very similar to those in Sec.\ref{11} and so we refrain from giving a detailed description. However, note that now the effect of anisotropy is made more conspicuous by the significant deviation of the curves from the corresponding isotropic ones. As in the previous section, here too, we observe the appearance of a minimal value of the dipole separation for the upper unstable branch arising out of the lower bound that was clamped upon $\tilde{K}$.

\section{Comparison among the different cases} \label{comp}
\begin{figure}[t]
\begin{center}
\subfigure[]{
\includegraphics[width=7.4cm,height=7.4cm, angle=-0]{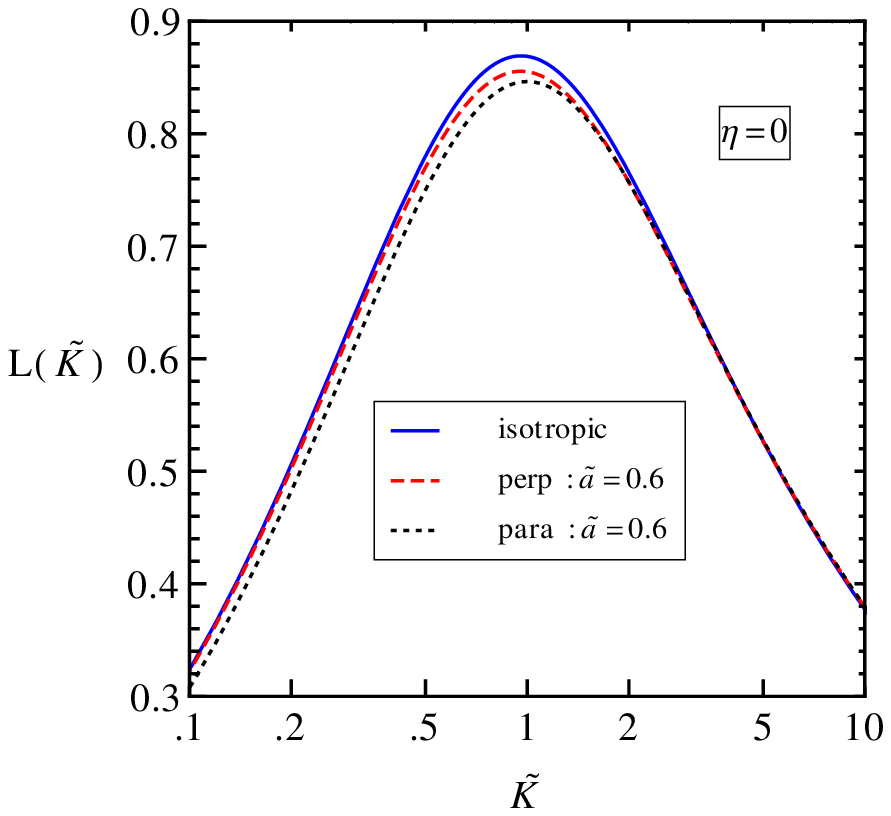}}
\hspace{0mm}
\subfigure[]{
\includegraphics[width=7.7cm,height=7.7cm, angle=-0]{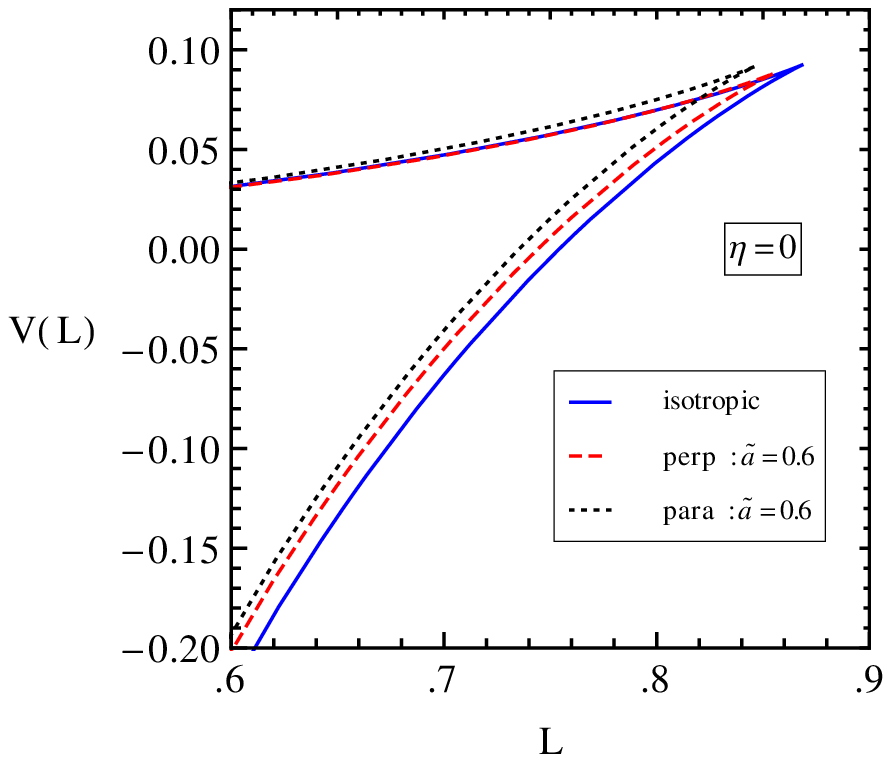}}
\caption{\label{cometa0} \small{(a) shows the plot of 
$L$ (normalized) as a function of $\tilde{K}$ with $\eta=0,\tilde{a}=0.6$ for
different orientations of the dipole and its direction of motion. (b) shows the plot
of properly normalized $V$ as a function of $L$ with
$\eta=0,\tilde{a}=0.6$ for the same set of orientations and direction of motion.}}
\end{center}
\end{figure}
\begin{figure}[t]
\begin{center}
\subfigure[]{
\includegraphics[width=7.4cm,height=7.4cm, angle=-0]{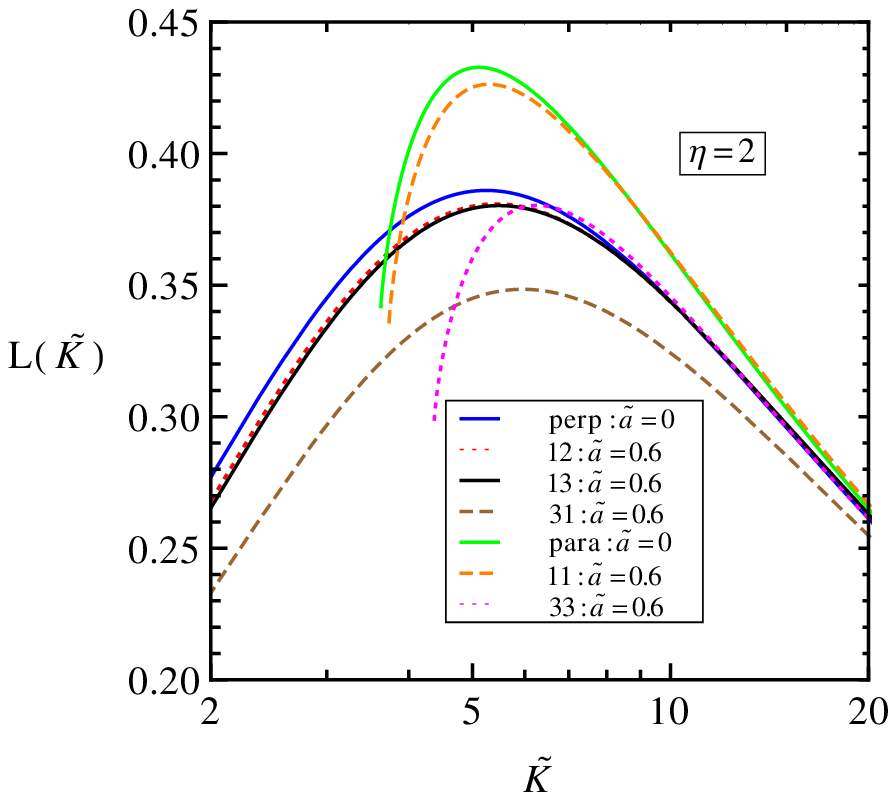}}
\hspace{0mm}
\subfigure[]{
\includegraphics[width=7.4cm,height=7.4cm, angle=-0]{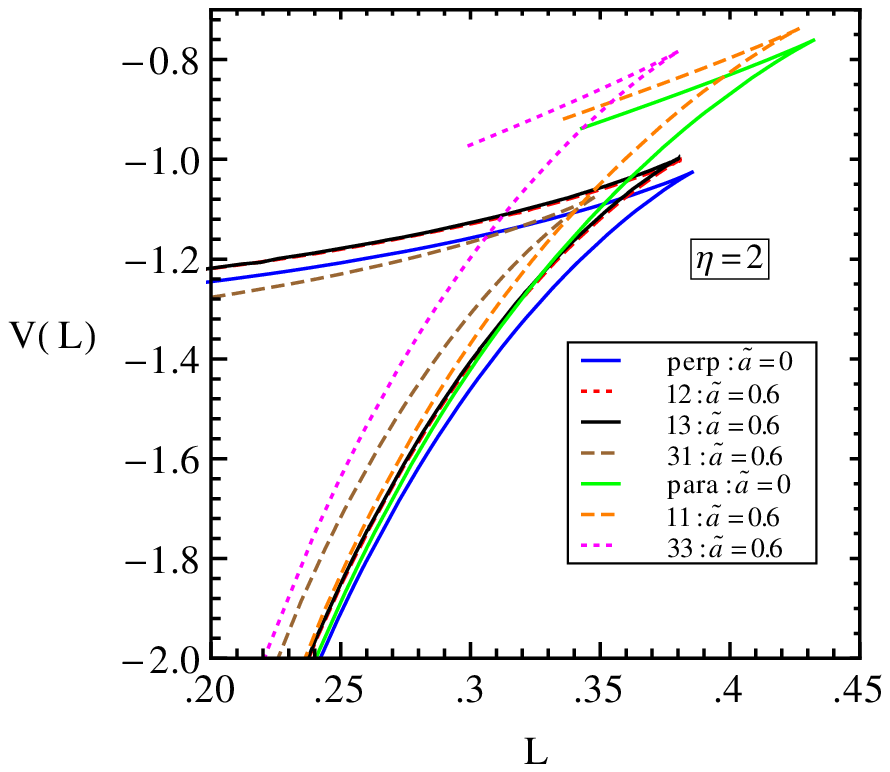}}
\caption{\label{cometa2} \small{(a) shows the plot of 
$L$ (normalized) as a function of $\tilde{K}$ with $\eta=2,\tilde{a}=0.6$ for
different orientations of the dipole and its direction of motion. (b) shows the plot
of properly normalized $V$ as a function of $L$ with
$\eta=2,\tilde{a}=0.6$ for the same set of orientations and direction of motion.}}
\end{center}
\end{figure}

In the previous section we have computed the \qq separation and the \qq potential for different orientations of the dipole and its velocity. Before concluding, let us do a comparative study of the effects of anisotropy in all the cases. In Fig.\ref{cometa0} we have given the $L(\tilde{K})$-$\tilde{K}$ and the $V(L)$-$L$ plots for the three surviving cases\footnote{ A little deliberation shows that in the static limit many of the cases collapse into each other and need not be considered separately.} for $\eta=0$ and $\tilde{a}=0.6$.  The legend in the figure needs a little explanation. While the blue line indicates the isotropic curve, `perp' indicates the dipole is moving in the transverse plane, perpendicular to the direction of anisotropy and `para' denotes the case where the dipole presents itself along the anisotropic direction.  While the presence of anisotropy makes itself felt in both the cases, the dipole is more affected when it is aligned parallel to the direction of anisotropy so that one can 
write, $L_{max}(para)<L_{max}(perp)<L_{max}(isotropic)$ and $V_{isotropic}<V_{perp}<V_{para}$. This observation corroborates the findings in \cite{Gia}. In Fig.\ref{cometa2} we have plotted the same quantities, now for $\eta=2$ and for all the configurations considered. Before delving into the details of the plots, let us again clarify the legend used. Note that now there are two isotropic plots, denoted by `perp' and `para' indicating the cases where the dipole lies perpendicular and parallel to the direction of motion respectively. $(ij)$ denotes the configuration where the dipole moves along $x^{i}$ and is aligned along $x^{j}$. Basically, one can distinguish between two sectors: one in which the dipole is perpendicular to its velocity (this contains `perp', (12), (13), (31)) and the one where it is parallel to its velocity (comprising of `para', (11), (33)). The general observation is that the screening length diminishes and the potential is weaker for all the cases $(ij)$ shown compared to the 
corresponding isotropic cases. The cases $(12)$ and $(31)$ which merged in the static case now splits up and we find that $(31)$ is severely affected when the combined effects of velocity and anisotropy are taken into account. This is evident both from the $L(\tilde{K})$-$\tilde{K}$ and the $V(L)$-$L$ plots. For this configuration $L_{max}$ drops drastically and also the rise in $V(L)$ is appreciable. As discussed earlier too, this is accounted for by the presence of the $\z{O}(y^{2})$ term in the anisotropic contribution, which makes the effect of anisotropy quite pronounced in this configuration. Both the $(13)$ and $(12)$ cases are mildly affected when effects of velocity and anisotropy act in conjunction. For these cases $L$ is slightly suppressed from the isotropic value whereas $V(L)$ registers a small increase. Turning to the other sector, we see that in $(11)$, $L$ decreases marginally which is accompanied by a corresponding small increase in the interaction potential when we introduce anisotropy and 
the velocity parameter together. However, the $(33)$ plots show a significant departure from the isotropic case. For the unstable high energy branch of the potential, the minimum allowed separation $L_{min}$ decreases in the order, $L_{min}(33)<L_{min}(11)<L_{min}(isotropic)$. On the whole, the plots suggest that the dipole separation and the potential are affected the most when the dipole moves along the anisotropic direction (both for the perpendicular and the parallel orientation and we hope, it will hold true for an other orientation in between these two extreme cases), and irrespective of the configuration, the presence of anisotropy makes the dipole more susceptible to dissociation. \\

\section{Discussion} \label{disc}
Before closing, it will be interesting to compare our observations with those extracted from other models of anisotropic plasma. In \cite{Stri} the heavy quark-antiquark static potential was computed in an anisotropic plasma employing the hard thermal loop approach. It was found out that the presence of anisotropy reduces the screening so that the potential, in general, gets strengthened and approaches the vacuum potential. The deviation from the isotropic screened potential increases as the value of the anisotropy parameter is increased. Further, the potential approaches the Coulomb potential faster when the dipole presents itself parallel to the direction of anisotropy. However, in the static case that we have studied here (and also in \cite{Gia}) introduction of anisotropy shifts the potential upwards, i.e. the potential now moves away from the screened one and even further away from the Coulomb potential. Here too, the dipole parallel to the direction of anisotropy gets more affected. Thus we find that 
our results (and those in \cite{Gia}) are qualitatively different from those obtained using field theoretic tools.
By introducing the velocity, we have shown here that for sufficiently large value of the velocity the effects of anisotropy on the dipole moving along the anisotropic direction will be the strongest. It might be interesting to attempt a similar study in the perturbative approach (such a study for the isotropic case has already been done in \cite{Mun}) and see if the introduction of the dipole velocity leads to results similar to that presented here. Another difference is that the results of \cite{Stri} hold for length scales $\sim \la_{D}=1/m_{D}$ where $m_{D}$ is the Debye mass. At this length scale the Coulomb part of the potential dominates over the linear part. However, our analysis here is not constrained in this aspect and in fact, in the static case we have identified a range $L_{p}<L<L_{max}$ where the string tension term dominates over the Coulomb one.  Of course, it will be naive on our part to read too much into these comparisons, since the physical models in the two cases are completely different,
 the primary difference being  that our analysis holds for strongly coupled theories whereas the perturbative calculations are valid only in the weak coupling limit.\\
Another key issue to address is the relevance of  the model we have considered here to real world QCD. This will determine whether the model can be reliably used to make predictions and/or explain RHIC and LHC results. 
One has to answer two questions - firstly, how do the anisotropic model examined here compare with hot Yang-Mills; secondly, how are the effects of anisotropy manifested in other models of anisotropic plasma, i.e., whether various models of anisotropic plasma share any generic feature and one is allowed to talk about a universality class. First of all, one has to keep in mind that, as stressed earlier, we have not yet found the gravity dual to QCD. While the attempt to obtain an exact gravity dual to QCD goes on, the best we can do at the moment is to propose gravity duals that come under the same universality class and hope that the qualitative behaviors of QCD are captured in these models. The deformed $\z{N}=4$ SYM that we have studied here shares many of its properties with the parent $\z{N}=4$ SYM and hence, a comparison of the deformed theory \textit{vis-a-vis} QCD can be done along the lines of SYM \textit{vis-a-vis} QCD except for some minor modifications arising out of anisotropy. Such a comparison 
for the isotropic case has been done in detail in \cite{Raj,Liu} and we present a brief discussion here following these papers. While the deformed $\z{N}=4$ SYM differs essentially from QCD on many counts, many of the differences cease to have much implications in the regime we have worked (the same holds for the isotropic case too). For example, $\z{N}=4$ SYM (and its deformed cousin) is a supersymmetric theory while QCD is not. However, the introduction of temperature ensures that the supersymmetry is explicitly broken so that it  is no longer an issue. Another point of difference often mentioned is that QCD is not a conformally invariant theory while $\z{N}=4$ SYM is. In this regard note that there are indications from lattice computations that QCD thermodynamics can well be considered as conformal in a temperature range $2T_{c}$ up to a upper limit not currently determined \cite{Liu}. In this regime, 
the deviation from conformality, which is measured by $\epsilon-3P$ ($\epsilon$ being the energy density), is not significant and SYM faithfully encapsulates the features of QCD. In fact, in the anisotropic case, the situation is somewhat better since the introduction of anisotropy renders the theory non-conformal. As can be found in the original papers \cite{Mat, Mat2}, the source of the conformal anomaly resides in the fact that the translational invariance along the radial direction is spoiled by the renormalisation of the on-shell supergravity action. A signature of this can be found in that the thermodynamic quantities (barring the entropy density) depend upon $a$ and $T$ separately rather than on the ratio $a/T$ \cite{Cher}. So it appears that the deformed version of SYM  may be a better approximation to QCD than its isotropic counterpart, particularly when the effects of anisotropy are significant, just after the plasma is produced. In view of these similarities it is not unnatural to presume that the 
two theories - deformed SYM  and QCD, show qualitatively similar behavior in many respect. However, to really arrive at a robust conclusion it is imperative that one first does similar computations with other models of anisotropic plasma and see if the anisotropic models themselves exhibit any universal features. This then brings us to the second question. However, in this regard we are handicapped by the fact that there are not many models of anisotropic plasma available for performing calculations. A comparison can be drawn with the results of \cite{Chak1} where, among other things, the screening length, the velocity-dependent \qq 
separation and the \qq potential were found out in thermal, non-commutative Yang-Mills plasma (NCYM). The presence of non-commutativity breaks the isotropy and this is reflected in the background metric, where the $x^{1}$-direction is taken to be the anisotropic one. Thus it is worthwhile to explore whether the NCYM can serve as a toy model to study the effects of anisotropy. The configuration considered in \cite{Chak1} corresponds to that in Sec.\ref{31} in the present paper. The counterpart of the anisotropy parameter is the non-commutativity parameter $\theta$ in \cite{Chak1}. Both the present work and \cite{Chak1} compute the screening length (albeit for different configurations) and it is tempting to compare the two and try to figure out a general scenario. While due caution should be exercised keeping in mind the different configurations, one finds that the expressions for $L_{max}$ are tantalisingly similar in both the cases.  To facilitate comparison, we give the result from \cite{Chak1} for the 
screening length  in NCYM when the non-commutativity is small,
\bea \label{LmaxNCYM}
L_{max}=\frac{1}{\sqrt{\pi}T}\frac{\G(3/4)}{\G(1/4)}\frac{2\sqrt{2}}{3^{3/4}}(1-v^{2})^{1/4}\left( 1-\frac{7}{2}\frac{\pi^{4}\hat{\lambda}T^{4}\theta^{2}}{(1-v^{2})}+...\right)
\eea
A comparison with (\ref{Lmax}) reveals that at the leading order (i.e., when the isotropic limit is recovered) the screening length scales as $(1-v^{2})^{1/4}$ in both the cases and also scales as $1/T$. At the subleading order (i.e., when the effects of anisotropy or non-commutativity make themselves manifest) we find that the correction to screening length scales with velocity as $(1-v^{2})^{-1}$ in either case. In the present case the correction term scales with anisotropy as $\tilde{a}^{2}$ whereas in the NCYM the correction term depends upon the non-commutative parameter as $\theta^{2}$. In both the cases, the screening length decreases as one increases $\tilde{a}$ or $\theta$. While the details are different and there are additional dependence upon the non-commutative coupling constant $\hat{\lambda}$ and temperature $T$ in (\ref{LmaxNCYM}) one observes that certain generic features are preserved. This points to the possibility that there may as well be some universality class under which both these 
models fall. This is reinforced by a comparison of the numerical results for the heavy quark potential in our case and that found in \cite{Chak1}. While in our case, the screening length falls  and the potential rises upwards with increasing the anisotropy, in \cite{Chak1} a rise in the non-commutativity parameter (but still keeping it small) leads to similar results. On a more general note, one can speculate that since our calculation hinges on the coupling of the string with the background metric, any source of anisotropy that leads to qualitatively similar background will result in 
almost similar sort of behavior for the heavy quark potential and the screening length independent of the exact details of the remaining supergravity field content.  In \cite{Chak1}, in addition, it was possible to study the effects of large value of the non-commutativity parameter. In the present paper, we have refrained from considering large values of anisotropy since our main motivation was to undertake an analytical study of the effects of anisotropy whereas to study effects of large anisotropy one would require to resort to numerical means from the very outset (for large values of anisotropy even the metric components are not known in an analytical form). Hence, our calculation is essentially a perturbative one when the deviation from the isotropic phase is small. Consequently, the results we have obtained are also qualitatively similar to that of the isotropic plasma. It will, of course, be interesting to investigate effects of large anisotropy as well for the different configurations considered here (
one such configuration was considered in \cite{Cher3}) and see how the results are affected. We have already seen that in the static case for small anisotropy the dipole oriented along the anisotropic direction is mostly affected whereas for large enough velocity the dipole moving along the anisotropic direction is mostly affected. In keeping with this, a very natural question to ask will be which one of the five configurations considered here will be most affected when $\tilde{a}$ too takes large values. \\
Another point we wish to emphasize is that the quark-antiquark pair we have considered is essentially infinitely massive. On the gravity side they were introduced through a fundamental string so that the probe quarks we are considering are in the fundamental representation. So if one wishes to relate our results to realistic scenarios of heavy ion collision, the analogue would be heavy quarkonium mesons. Thus our calculations suggest that in the initial stages of QGP (but after equilibriation), presence of anisotropy leads to an enhancement of quarkonium dissociation whose direct fallout will be an increase in the suppression of quark-antiquark bound states like $J/\Psi$. Another very relevant point worth discussing is how the presence of anisotropy affects energy loss of the probe quarks. There are primarily two modes of energy loss: one is collisional which is measured by the drag force experienced by the probe; another is radiative which is quantified by the jet quenching parameter $\hat{q}$. The jet 
quenching parameter 
can be readily obtained from the Wilson loops we have computed by taking the $\eta \rightarrow \infty$ limit followed by taking the boundary to infinity. However, $\hat{q}$ has been calculated \cite{Cher2} (using the light-cone coordinates) for the most general orientation and arbitrary value of the anisotropy parameter. It was observed that $\hat{q}$ depends upon the relative orientation of the anisotropic direction, direction of the dipole and that along which the momentum broadening is measured. Further, its value can be larger or smaller than the isotropic case depending upon whether the comparison is made at equal temperature or equal entropy density.
The drag force has been obtained in \cite{Cher} and it was found that the drag coefficient can be smaller or larger than its isotropic counterpart depending upon the velocity and the direction of motion. We refer the reader to \cite{Cher,Cher2} for a more detailed discussion regarding the energy loss of probe quarks in hot anisotropic QGP.
\\

\section{Conclusion} \label{conc}
Finally, we conclude with a brief summary of the results obtained in the paper. We have found out the velocity-dependent \qq separation and the \qq potential in a strongly coupled anisotropic plasma at finite temperature via the  gauge/gravity duality. The gauge theory we take is a  deformation of the $\z{N}=4$ SYM and we take the gravity dual as proposed in \cite{Mat,Mat2}. Barring the screening length in a special case, in all the other cases we presented numerical results. The general observation is that when we turn on a small value of the anisotropy parameter, the screening length ($L_{max})$ decreases and the \qq interaction becomes weaker so that the dipole becomes more prone to dissociation. We considered five different cases, depending upon the direction of motion of the dipole and the direction along which it is aligned. While the generic features of the plots are the same in all the cases, the minute details vary from case to case. In particular, when the dipole lies along the direction of 
anisotropy the effects are manifested more prominently in the static case. However, for finite velocity, it is the dipole moving along the anisotropic direction that is affected the most. We set the rapidity parameter $\eta=0$ and recover the static \qq separation  and the static \qq potential. In these cases, our findings are consistent with those recently obtained in \cite{Gia}.  Finally, we also compared the results obtained in this model \textit{vis-a-vis} some other models. In particular, we found that the results for the static dipole potential provided here (and also in \cite{Gia}) are different from those obtained using standard perturbative field-theoretic techniques in the weakly coupled regime. On the other hand, all our results are remarkably similar with those obtained for hot non-commutative Yang-Mills theory where the presence of non-commutativity can be seen as a source of anisotropy.

\section*{Acknowledgements}
The authors would like to thank  Shibaji Roy and Munshi G. Mustafa for various fruitful discussions and the anonymous referee whose suggestions, we hope, have helped us improve the paper.

\end{document}